\documentclass{aa}    
\usepackage{txfonts}
\usepackage{longtable}  
\usepackage{natbib,twoopt}
\usepackage{graphicx}

\bibpunct[, ]{(}{)}{,}{a}{}{,}

\usepackage{color}

\newcommand{\Teff}{T_{\rm eff}}
\newcommand{\eps}[1]{\log\varepsilon_{\rm #1}}
\newcommand{\kms}{km\,s$^{-1}$}
\newcommand{\iso}[2]{\mbox{$^{#1}{\rm #2}$}}
\newcommand{\eu}[5]{\mbox{$#1\,^#2{\rm #3}^{#4}_{\rm #5}$}}
\newcommand{\kH}{$S_{\!\!\rm H}$}    
\newcommand{\Eexc}{$E_{\rm exc}$}
\newcommand{\Vmic}{V_{\rm mic}}

\begin{document}

\title{Influence of inelastic collisions with hydrogen atoms on the non-LTE modelling of \ion{Ca}{i} and \ion{Ca}{ii} lines in late-type stars}

\author{
  L. Mashonkina\inst{1,2} \and
  T. Sitnova\inst{2} \and
  A. K. Belyaev\inst{3}  
}
 
\offprints{L. Mashonkina; \email{lima@inasan.ru}}
\institute{
     Universit\"ats-Sternwarte M\"unchen, Scheinerstr. 1, D-81679 M\"unchen, 
     Germany \\ \email{lyuda@usm.lmu.de}
\and Institute of Astronomy, Russian Academy of Sciences, Pyatnitskaya st. 48, RU-119017 Moscow, 
     Russia \\ \email{lima@inasan.ru}
\and Herzen University, Moika 48, RU-191186 St. Petersburg, Russia
}

\date{Received   /  Accepted }

\abstract
{
We performed the non-local thermodynamic equilibrium (non-LTE, NLTE) calculations for \ion{Ca}{i-ii} with the updated model atom that includes new quantum-mechanical rate coefficients for \ion{Ca}{i} + \ion{H}{i} collisions from two recent studies
and investigated the accuracy of calcium abundance determinations using the Sun, Procyon, and five metal-poor (MP,  $-2.6 \le\ $[Fe/H] $ \le -1.3$) stars with  well-determined stellar parameters.
Including \ion{H}{i} collisions substantially reduces over-ionisation of \ion{Ca}{i} in the line formation layers compared with the case of pure electronic collisions and thus the NLTE effects on  abundances derived from \ion{Ca}{i} lines. We show that both collisional recipes lead to very similar NLTE results.
As for \ion{Ca}{ii}, the classical Drawinian rates scaled by \kH\ = 0.1 are still applied. 
When using the subordinate lines of \ion{Ca}{i} and the high-excitation lines of \ion{Ca}{ii},
NLTE provides the smaller line-to-line scatter compared with the LTE case for each star.
For Procyon, NLTE removes a steep trend with line strength among strong \ion{Ca}{i} lines seen in LTE and leads to consistent [Ca/H] abundances from the two ionisation stages. 
In the MP stars, the NLTE abundance from \ion{Ca}{ii} 8498\,\AA\ agrees well with the abundance from the \ion{Ca}{i} subordinate lines, in contrast to LTE, where the abundance difference grows towards lower metallicity and reaches 0.46~dex in BD~$-13^\circ$3442 ([Fe/H] = $-2.62$).
NLTE largely removes abundance discrepancies between the high-excitation lines of \ion{Ca}{ii} and \ion{Ca}{ii} 8498\,\AA\ obtained for our four [Fe/H] $< -2$ stars under the LTE assumption.
We investigated the formation of the \ion{Ca}{i} resonance line in the [Fe/H] $< -2$ stars. When the calcium abundance varies between [Ca/H] $\simeq -1.8$ and $-2.3$, photon loss in the resonance line itself in the uppermost atmospheric layers drives the strengthening of the line core compared with the LTE case, and this effect prevails over the weakening of the line wings, resulting in negative NLTE abundance correction and underestimation of the abundance derived from \ion{Ca}{i} 4226\,\AA\ compared with that from the subordinate lines, by 0.08 to 0.32~dex. This problem may be related to the use of classical homogeneous (1D) model atmospheres. The situation is improved when the calcium abundance decreases and the \ion{Ca}{i} 4226\,\AA\ line formation depths are shifted into deep atmospheric layers that are dominated by over-ionisation of \ion{Ca}{i}. However, the departures from LTE are still underestimated for \ion{Ca}{i} 4226\,\AA\ at [Ca/H] $\simeq -4.4$  (HE~0557-4840). Consistent NLTE abundances from the \ion{Ca}{i} resonance line and the \ion{Ca}{ii} lines are found for HE~0107-5240 and HE~1327-2326  with [Ca/H] $\le -5$. 
Thus, the \ion{Ca}{i}/\ion{Ca}{ii} ionisation equilibrium method can successfully be applied to determine surface gravities of [Ca/H] $\precsim -5$ stars.
We provide the NLTE abundance corrections for 28 lines of \ion{Ca}{i} in a grid of model atmospheres  with 5000 K $\le \Teff\ \le$ 6500 K,  2.5 $\le$ log~g $\le$ 4.5,  --4 $\le$ [Fe/H] $\le$ 0, which is suitable for abundance analysis of FGK-type dwarfs and subgiants.
}
\titlerunning{Non-LTE line formation for \ion{Ca}{i-ii} in late-type stars}
\authorrunning{Mashonkina et al.}
\maketitle

\section{Introduction}\label{Sect:intro}

Calcium plays an important role in studies of late-type stars. The subordinate lines of neutral Ca are suitable for spectroscopic analysis over a wide range of Ca abundance from super-solar values down to [Ca/H]\footnote{In the classical notation, where [X/H] = $\log(N_{\rm X}/N_{\rm
    H})_{star} - \log(N_{\rm X}/N_{\rm H})_{Sun}$.} = $-3.5$ \citep{Cohen2013}.
Calcium is the only chemical element that is visible in the two ionisation stages in the ultra metal-poor (UMP, [Fe/H] $< -4$) and hyper metal-poor (HMP, [Fe/H] $< -5$) stars \citep{2002Natur.419..904C,Frebeletal:2005,2015ApJ...810L..27F,Norrisetal:2007,2011Natur.477...67C}, and the \ion{Ca}{i} and \ion{Ca}{ii} lines can be potent tools in determining accurate stellar surface gravity \citep[see, for example,][]{2009ApJ...698..410K} and the Ca abundance itself. 
The \ion{Ca}{ii} H\&K resonance lines remain measurable in spectra of the most iron-poor stars in which no iron line can be detected, for example, SMSS\,0313-6708, with [Ca/H] = $-7$ \citep{2014Natur.506..463K}, and SDSS\,J1035+0641, with [Ca/H] = $-5$ \citep{2015A&A...579A..28B}, and in this way, they provide an opportunity to examine the earliest stage of $\alpha$-process nucleosynthesis in our Galaxy. 
The subordinate lines of ionised Ca at 8498, 8542, and 8662\,\AA\ are among the strongest features in the near-infrared (IR) spectra of FGK-type stars with metallicity down to [Fe/H] = $-5$ \citep{2012A&A...542A..51C} and can serve as powerful abundance and metallicity indicators for distant objects in our Galaxy and nearby galaxies via medium-resolution spectroscopy. These lines lie at the focus of large spectroscopic surveys of the Milky Way, such as the Radial Velocity Experiment \citep[RAVE][]{RAVE2006AJ....132.1645S} and the ESA Gaia satellite mission \citep{2001A&A...369..339P}, and surveys of nearby galaxies like the CaT (\ion{Ca}{ii} IR triplet lines) survey \citep{2008ApJ...681L..13B}.

In stellar atmospheres with $\Teff > 4500$~K, neutral calcium is a minority species, and its statistical equilibrium (SE) can easily deviate from thermodynamic equilibrium owing to deviations of the mean intensity of ionising radiation from the Planck function. Early investigations of the SE of \ion{Ca}{i} in the Sun and Procyon \citep{1985A&A...149...21W} and the models representing atmospheres of FGK-type stars \citep{1991MNRAS.251..369D} found that the over-ionisation effects lead to depleted level populations for \ion{Ca}{i} compared with the thermodynamical equilibrium populations. 
\citet{Thevenin2000} considered the non-local thermodynamical equilibrium (NLTE) line formation for \ion{Ca}{i} to determine the Ca abundances of an extended sample of cool stars with various metallicities. NLTE calculations of the \ion{Ca}{ii} IR triplet lines in the moderately metal-poor model atmospheres were performed by \citet{1992A&A...254..258J} and \citet{2005A&A...430..669A}.

A comprehensive model atom for calcium was built by \citet{mash_ca} in order to consider the NLTE line formation for \ion{Ca}{i} and \ion{Ca}{ii} through a wide range of spectral types when the Ca abundance varies from the solar value down to [Ca/H] = $-5$. The treated NLTE method was applied by \citet{Norrisetal:2007} and \citet{2009ApJ...698..410K} to constrain surface gravity of an UMP star HE~0557-4840 and a HMP star HE~1327-2326, respectively. Using the same method, \citet{lick_paperII} established that the [Ca/Fe] ratios of a sample of the $-2.62 \le {\rm [Fe/H]} \le +0.24$ dwarf stars form a metal-poor plateau at a height of 0.3~dex that is similar to plateaus for the other $\alpha$-process elements Mg, Si, and Ti, and the knee occurs at common [Fe/H] $\simeq -0.8$.
Based on NLTE calculations of \citet{mash_ca}, \citet{Starkenburg2010} proposed to revise the CaT~--~[Fe/H] relation in the low-metallicity regime by taking substantial departures from LTE for the \ion{Ca}{ii} IR triplet lines into account. 

 The NLTE method was treated by
\citet{2011MNRAS.418..863M} to evaluate the NLTE effects for the infrared \ion{Ca}{i} and \ion{Ca}{ii} lines 
in the grid of models representing the atmospheres of cool giants with metallicity down to [Fe/H] = $-4$. \citet{2012A&A...541A.143S} performed the NLTE abundance determinations from lines of \ion{Ca}{i} and \ion{Ca}{ii} in a sample of very metal-poor (VMP) cool giants from the Large Programme First Stars \citep{Cayrel2004}. 

The need for a new NLTE analysis of \ion{Ca}{i-ii} is motivated by the recent quantum-mechanical calculations of \citet{Belyaev2016_Ca}, \citet{2016PhRvA..93d2705B}, and \citet[][hereafter, MGB17]{ca1_hydrogen} 
for inelastic \ion{Ca}{i}+\ion{H}{i} collisions. In all the previous NLTE studies, \ion{H}{i} collision rates were calculated using the rough theoretical approximation of \citet{Drawin1968,Drawin1969}, as implemented by \citet{Steenbock1984}, and they were scaled by a factor \kH, which was constrained empirically. For example, \citet{mash_ca}  estimated \kH\ = 0.1 from inspection of different influences of \ion{Ca}{i}+\ion{H}{i} collisions on the \ion{Ca}{i} and \ion{Ca}{ii} lines in the nine reference stars with well-determined stellar parameters and high-quality observed spectra. Accurate quantum-mechanical rate coefficients,
namely, those from \citet{Belyaev2016_Ca}, were only applied by \citet{Mashonkina_dnlte2016} to compute the NLTE abundance corrections for lines of \ion{Ca}{i} in the cool giant model atmospheres. 

By applying the most up-to-date atomic data available so far, this study aims to investigate the accuracy of calcium abundance determinations using
the Sun, Procyon, and a number of VMP, UMP, and HMP stars and to provide the users with the NLTE abundance corrections for \ion{Ca}{i} lines in the models representing atmospheres of FGK-type stars. The model atom from \citet{mash_ca} is taken as a basic model.

The paper is organised as follows. Section\,\ref{Sect:NLTE} describes the method of NLTE calculations for \ion{Ca}{i-ii}. Observations and atmospheric parameters of our sample of stars are reviewed in Sect.\,\ref{Sect:basics}. In Section\,\ref{Sect:stars1} we determine abundances from lines of \ion{Ca}{i} and \ion{Ca}{ii} in the Sun and the sample stars and investigate whether applying accurate \ion{H}{i} collision rates leads to agreement between the two ionisation stages.
In Section\,\ref{Sect:ump} we revise Ca abundances of the HMP and UMP stars, HE~0107-5240 \citep[${\rm [Fe/H]} \simeq -5.3$][]{HE0107_ApJ}, HE~0557-4840 \citep[${\rm [Fe/H]} \simeq -4.8$][]{Norrisetal:2007}, and HE~1327-2326 \citep[${\rm [Fe/H]} \simeq -5.5$][]{Aoki_he1327} and test the NLTE modelling of the \ion{Ca}{i} resonance line in the VMP atmospheres. 
The NLTE abundance corrections for lines of \ion{Ca}{i} in a grid of models representing the atmospheres of FGK-type stars are predicted in Sect.\,\ref{sect:corrections}. 
Our recommendations and conclusions are given in Sect.\,\ref{conclusion}.

\section{Method of NLTE calculations for \ion{Ca}{i-ii}} \label{Sect:NLTE}

In this section, we describe updates of the model atom \ion{Ca}{i-ii} and inspect the influence of inelastic collisions with \ion{H}{i} on the SE of calcium. The coupled radiative transfer and SE equations are solved with the {\sc DETAIL} code
\citep{detail} based on the accelerated lambda iteration method 
\citep[recipe of][]{rh91,rh92}. The opacity package in {\sc DETAIL} was updated as described by \citet{mash_fe}. This research uses the MARCS homogeneous plane-parallel model atmospheres with standard abundances \citep{Gustafssonetal:2008} available on the MARCS web site\footnote{\tt
  http://marcs.astro.uu.se}. They were interpolated at the necessary $\Teff$,
log~g, and iron abundance [Fe/H], using the FORTRAN-based routine written by Thomas
Masseron that is available on the same website.

\subsection{Updated model atom} \label{Sect:model}

\begin{figure*}  

\includegraphics[width=80mm]{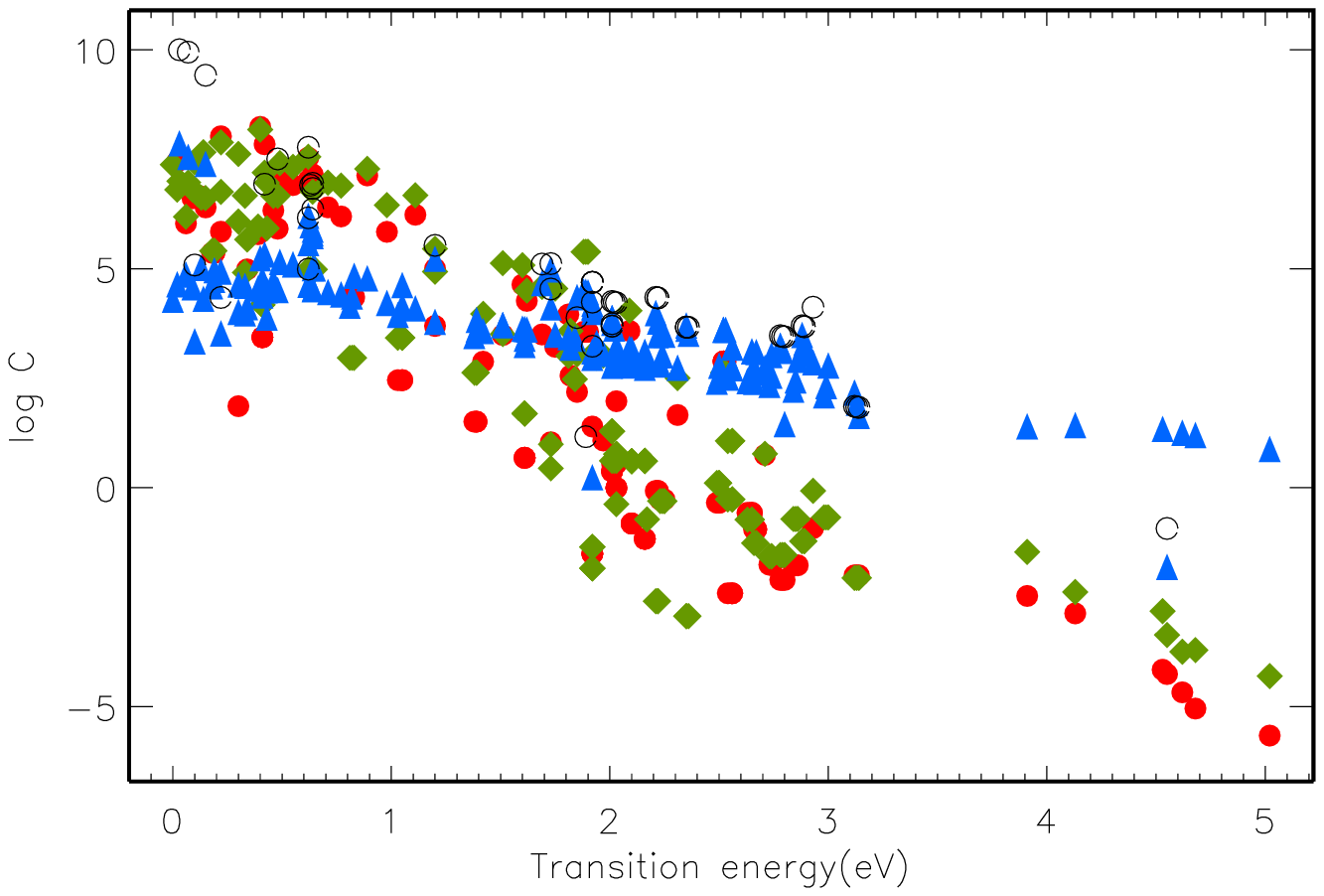}
\includegraphics[width=80mm]{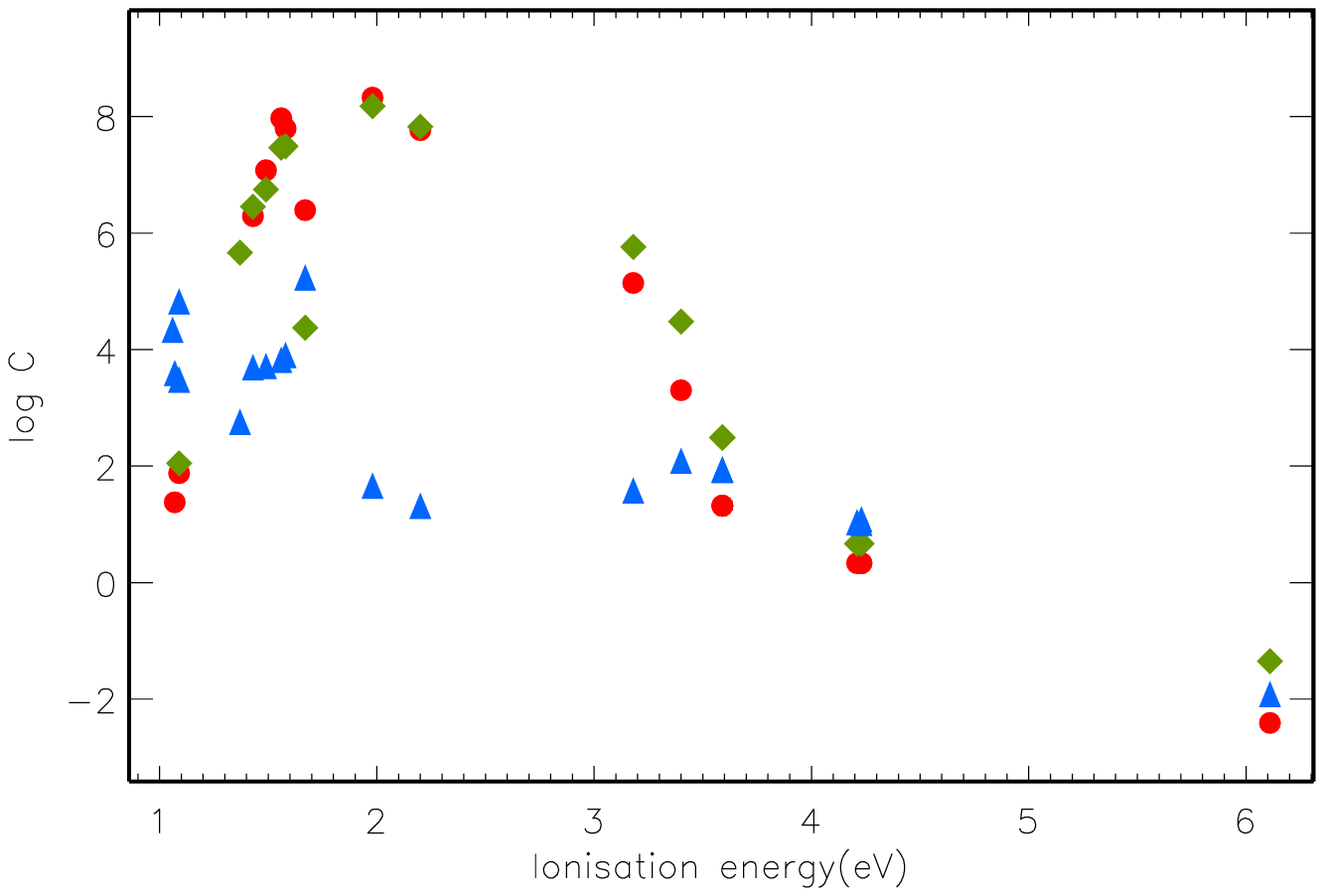}
\caption{Left panel: \ion{Ca}{i} excitation rates (in s$^{-1}$), log~C, for electron impact (triangles) compared with the rates for \ion{H}{i} collisions from quantum-mechanical calculations of MGB17 (filled circles) and B17u (rhombi) and compared with the scaled (\kH\ = 0.1) Drawinian rates (open circles). Right panel: rates, log~C, of the processes \ion{Ca}{i} + ${\rm e^-} \rightarrow$ \ion{Ca}{ii} + $2{\rm e^-}$ and \ion{Ca}{i} + \ion{H}{i} $\rightarrow$ \ion{Ca}{ii} + ${\rm H}^-$ using similar symbols. The calculations were made with $T = 5340$~K, log~$N_{\rm e}$(cm$^{-3}$) = 12.5, and log~$N_{\rm H}$(cm$^{-3}$) = 17.}
\label{fig:rates}
\end{figure*}

This study applies the NLTE method developed by \citet[][hereafter, Paper~I]{mash_ca} with the modifications concerning collisional rate computations. We briefly describe the atomic data we used.
The model atom includes 63 levels of \ion{Ca}{i}, 37 levels of \ion{Ca}{ii}, and the ground state of \ion{Ca}{iii}. For radiative transitions, we used accurate data on
photoionisation cross-sections and transition probabilities from the Opacity
Project \citep[OP; see][for a general review]{1994MNRAS.266..805S}, which are accessible in the TOPBASE\footnote{http://cdsweb.u-strasbg.fr/topbase/topbase.html} database. In the SE calculations, inelastic collisions with electrons and hydrogen atoms leading to both excitation and ionisation are taken into account. For electron impact excitation, detailed results from the $R-$matrix calculations are available for ten transitions from the ground state in \ion{Ca}{i} \citep{2001ADNDT..77...87S} and all the transitions between levels of the $n \le 8$ configurations of \ion{Ca}{ii} \citep{ca2_bautista}. For the remaining bound-bound transitions, approximate formulae were used, namely, the impact parameter method \citep[IPM,][]{1962PPS....79.1105S} for the allowed transitions and a collision strength of 1.0 and 2.0 for the optically forbidden transitions of \ion{Ca}{i} and \ion{Ca}{ii}, respectively. Electron impact ionisation cross-sections were calculated by applying the formula of \citet{1962amp..conf..375S} with threshold photoionisation
cross-sections from the OP data.

A novelty of this research is that we account for 
the ion-pair production from the ground and excited states of \ion{Ca}{i} and mutual neutralisation (charge-exchange reactions),

\ion{Ca}{i}($nl$) + \ion{H}{i} $\leftrightarrow$ \ion{Ca}{ii}(\eu{4s}{2}{S}{}{}) + H$^-$

\noindent
and \ion{H}{i} impact excitation and de-excitation processes in \ion{Ca}{i}, with the rate coefficients from quantum-mechanical calculations. The required data were taken from two different studies. 
\citet{2016PhRvA..93d2705B} developed a theoretical method for estimating cross-sections and rates for excitation and charge-transfer processes in low-energy hydrogen-atom collisions with neutral atoms, based on an asymptotic two-electron linear combination of atomic orbitals (LCAO) model of ionic-covalent interactions in the neutral atom-hydrogen-atom system and the multichannel Landau-Zener model.
The rate coefficients used in our study are updated data, including a corrected H$^-$ wavefunction. Hereafter, we refer to these data as B17u.
The rate coefficients of \citet{ca1_hydrogen} come from the accurate highly correlated ab initio electronic structure calculations followed by both the multichannel Landau-Zener model and the probability current method for nuclear dynamical calculations.

Figure~\ref{fig:rates} displays the excitation rates depending on the transition energy, $E_{lu}$, and the ion-pair production rates depending on the level ionisation energy, $\chi_l$, computed with the rate coefficients from two sources, MGB17 and B17u. Here, the data correspond to a kinetic temperature of $T = 5340$~K and an \ion{H}{i} number density of log~$N_{\rm H}$(cm$^{-3}$) = 17 that are characteristic of the line-formation layers (log~$\tau_{5000} = -0.52$) in the model atmosphere with $\Teff$ / log~g / [Fe/H] = 5780~K / 3.70 / $-2.46$. It is evident that for any common transition, both sources provide very similar collisional rates.
Collisions with \ion{H}{i} are more efficient compared with electron impacts in exciting the $E_{lu} \precsim$ 1.2~eV transitions, and the ion-pair production rates are 
substantially higher than the electron-impact ionisation rates for atomic levels with $\chi_l <$ 3.3~eV.
In the next subsection, we inspect the influence of employing both MGB17 and B17u rate coefficients on the SE calculations and the calcium abundance determinations. The NLTE results in Sects.\,\ref{Sect:stars1}-\ref{sect:corrections} are based on using the MGB17 data.

Collisions of \ion{Ca}{ii} with \ion{H}{i} were treated using the Drawinian rates that are scaled by a factor of \kH\ = 0.1, as recommended in Paper~I.
We neglected collisions with \ion{H}{i} for the \ion{Ca}{i} transitions for which \citet{2016PhRvA..93d2705B} and \citet{ca1_hydrogen} did not provide rate coefficients, and also for the forbidden transitions in \ion{Ca}{ii}.


\subsection{Influence of inelastic collisions with \ion{H}{i} on the statistical equilibrium of \ion{Ca}{i} and calcium abundance determinations}\label{Sect:effect}

\begin{figure*}
\resizebox{88mm}{!}{\includegraphics{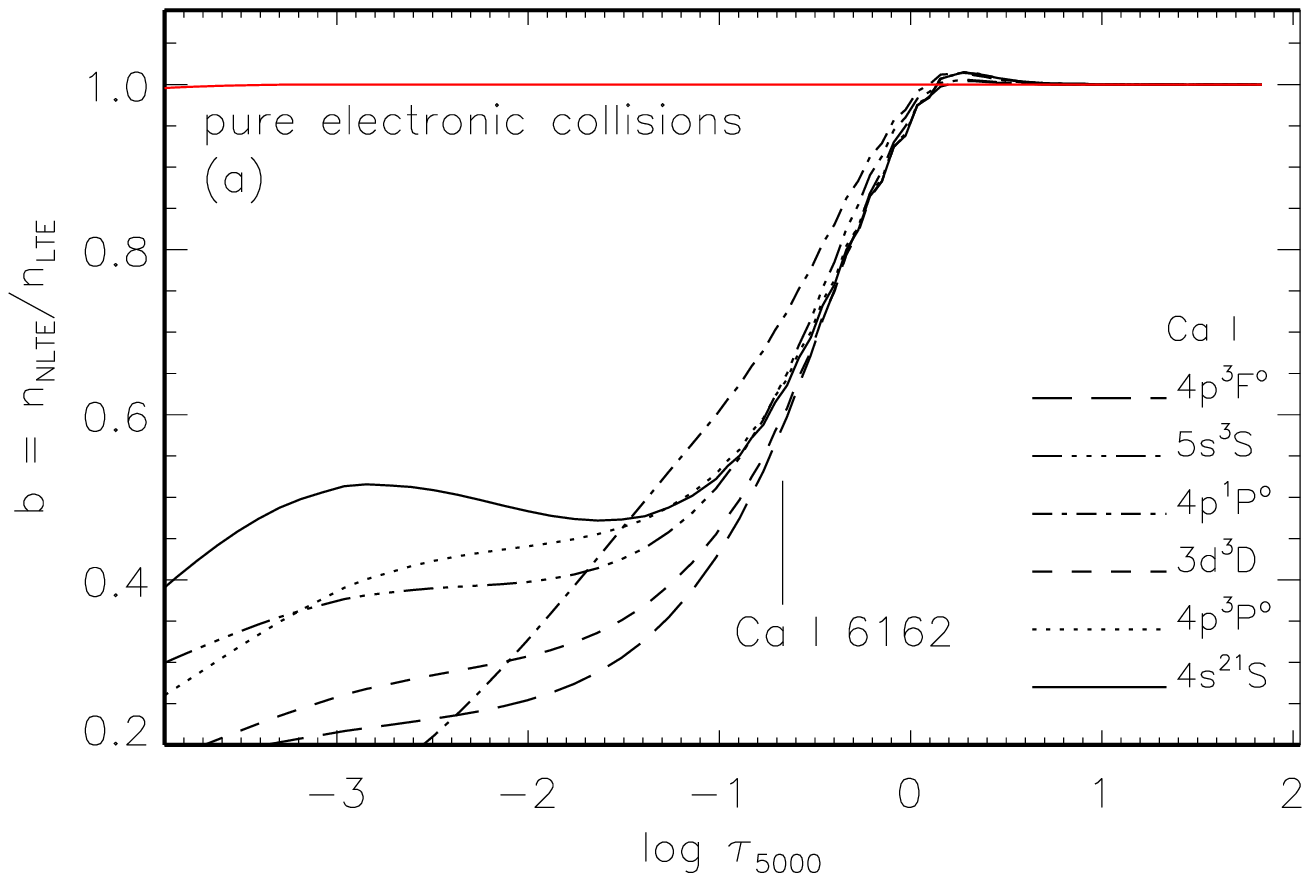}}
\resizebox{88mm}{!}{\includegraphics{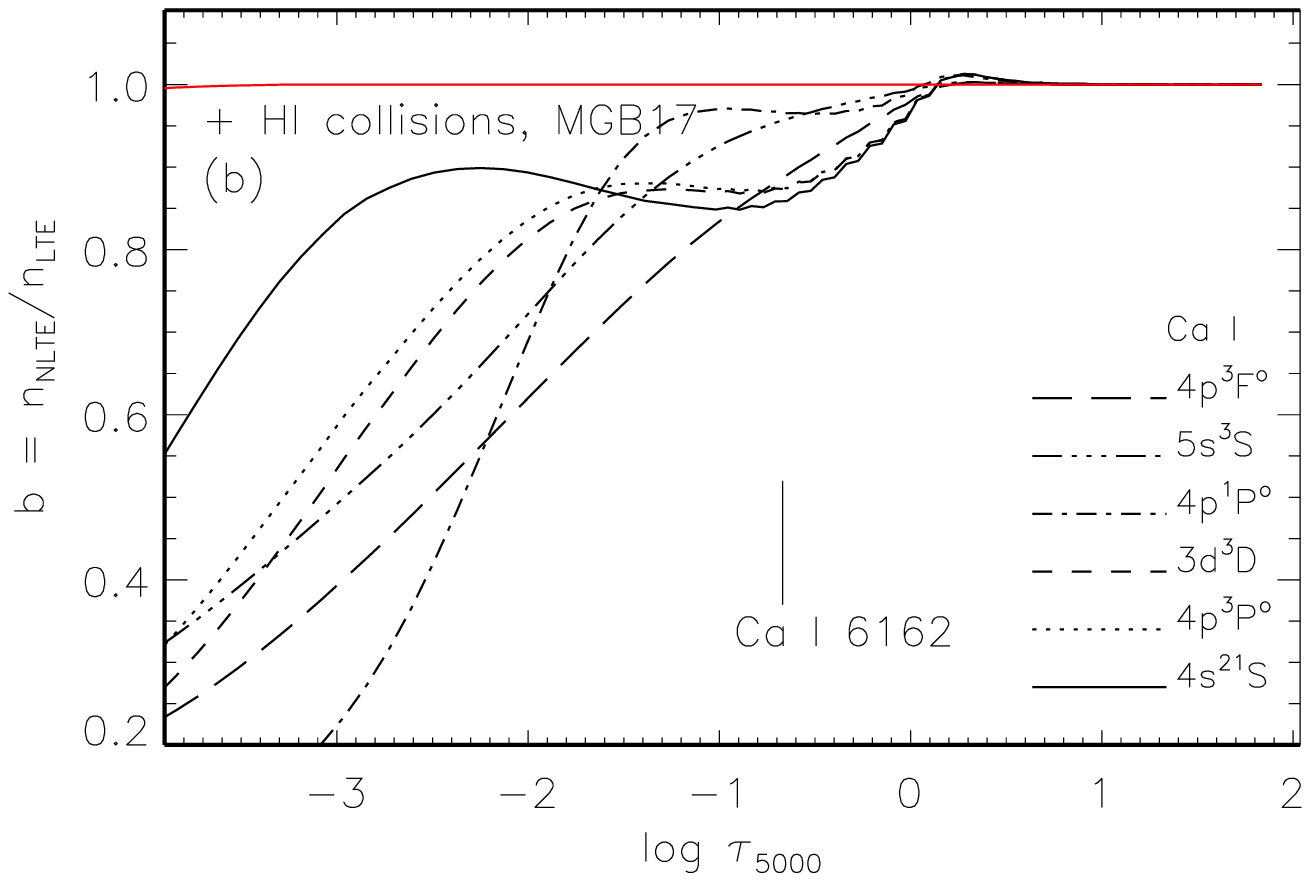}}

\resizebox{88mm}{!}{\includegraphics{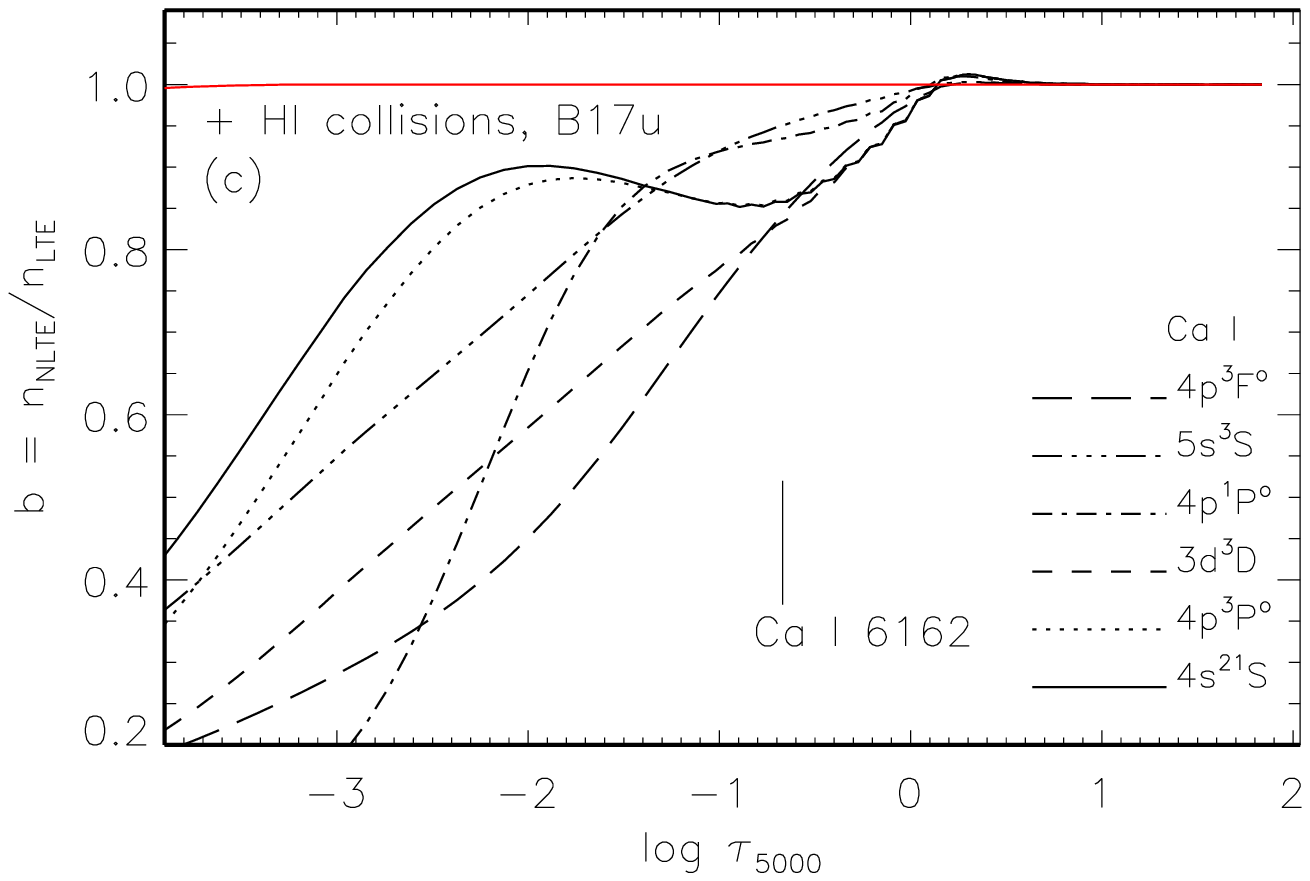}}
\resizebox{88mm}{!}{\includegraphics{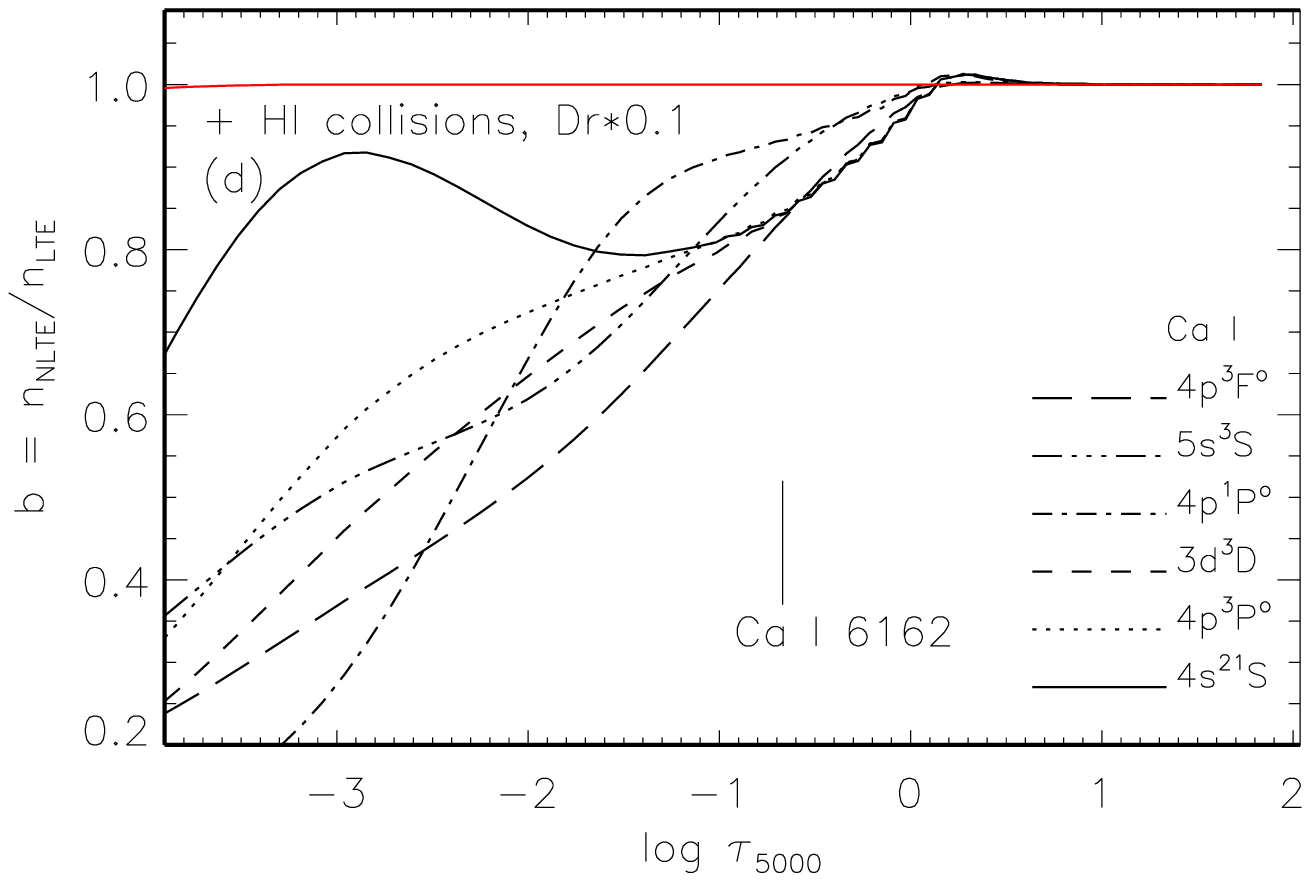}}
\caption[]{Departure coefficients, b, for selected levels of \ion{Ca}{i} from calculations using different treatment of \ion{H}{i} collisions: (a) - no \ion{H}{i} collisions, that is, pure electronic collisions; (b) - MGB17; (c) - B17u; (d) - Drawinian rates scaled by \kH\ = 0.1. 
For the ground state of \ion{Ca}{ii}, b = 1 (red continuous line) throughout the atmosphere. In each panel, the tick mark indicates the location of line centre optical depth unity for \ion{Ca}{i} 6162\,\AA. Here, all the calculations were performed with a common model atmosphere: 5780/3.70/$-2.46$ and [Ca/Fe] = 0.4.} \label{Fig:bf_ca1}
\end{figure*}

We chose the VMP model atmosphere, namely, 5780/3.70/$-2.46$, to investigate the influence of inelastic collisions with \ion{H}{i} and their different treatment on the NLTE results for \ion{Ca}{i}. 
The calculations were performed for five different line-formation scenarios based on pure electronic collisions (a), including \ion{H}{i} collisions with the rate coefficients from MGB17 (b) and B17u (c), 
removing all the b-b transitions (e) in the MGB17 recipe, and for comparison, using the Drawinian rates scaled by \kH\ = 0.1 (d).
Figure\,\ref{Fig:bf_ca1} displays the departure coefficients, ${\rm b = n_{NLTE}/n_{LTE}}$, for selected levels of \ion{Ca}{i} in the scenarios (a)-(d). Here,
${\rm n_{NLTE}}$ and ${\rm n_{LTE}}$ are the statistical equilibrium and thermal (Saha-Boltzmann) number densities, respectively. It is worth noting that each of the triplet terms, \eu{4p}{3}{P}{\circ}{} and \eu{3d}{3}{D}{}{}, is displayed as single level because the fine-splitting levels, which are treated explicitly in the model atom, reveal very similar departure coefficients.
Since \ion{Ca}{ii} is the majority species, variations in the treatment of \ion{Ca}{i} + \ion{H}{i} collisions nearly do not affect populations of the \ion{Ca}{ii} levels. Therefore, only the ground state of \ion{Ca}{ii} is plotted in Fig.\,\ref{Fig:bf_ca1}.

As shown in the earlier NLTE studies \citep[see][and references therein]{mash_ca}, the main NLTE mechanism for \ion{Ca}{i} in the stellar parameter range, with which we are concerned here, is the ultra-violet (UV) over-ionisation caused by superthermal radiation of a non-local origin below the thresholds of the low excitation levels. We obtain that all levels of \ion{Ca}{i} are underpopulated above $\log \tau_{5000} = 0$ in the 5780/3.70/$-2.46$ model atmosphere, independent of the treatment of collisional rates. As expected, the departures from LTE are smaller when collisions with \ion{H}{i} are included. For example, in the atmospheric layers, where the line core of \ion{Ca}{i} 6162\,\AA\ forms, at $\log \tau_{5000} = -0.7$, the departure coefficients of its lower (\eu{4p}{3}{P}{\circ}{2}) and upper (\eu{5s}{3}{S}{}{1}) levels amount to b$_l$ = 0.62 and b$_u$ = 0.62 in case of pure electronic collisions, while they are closer to unity in the MGB17 and B17u scenarios: b$_l$ = 0.875, b$_u$ = 0.954 and b$_l$ = 0.859, b$_u$ = 0.948, respectively. 
We find that the charge-exchange reactions influence the SE of \ion{Ca}{i} to a greater extent than the \ion{H}{i}-impact excitation. When we
removed all the bound-bound (b-b) transitions in the MGB17 scenario, the 
departure coefficients changed only slightly. For example, we obtain b(\eu{4p}{3}{P}{\circ}{}) = 0.862 and b(\eu{5s}{3}{S}{}{1}) = 0.942 at $\log \tau_{5000} = -0.7$.


It is worth noting that although the \citet{Drawin1968,Drawin1969} formalism does not
provide a realistic description of the physics involved, using the scaled Drawinian rates catches a main part of the effect of collisions with \ion{H}{i} on the SE of \ion{Ca}{i}. The b$_l$ = 0.845 and b$_u$ = 0.900 obtained for this option at $\log \tau_{5000} = -0.7$ are indeed closer to the MGB17 and B17u scenarios than to the case of pure electronic collisions. 

\begin{figure}
\resizebox{88mm}{!}{\includegraphics{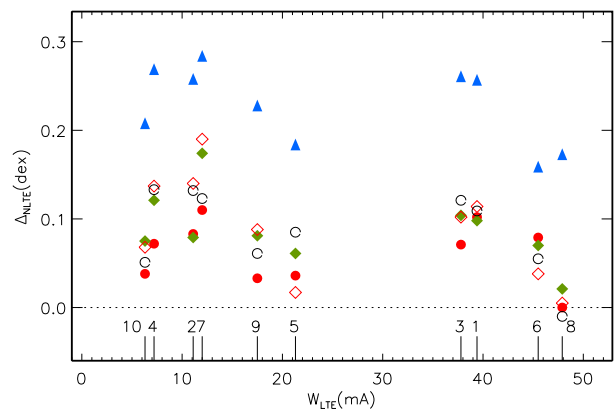}}
\caption[]{NLTE abundance corrections for the lines of \ion{Ca}{i} in the 5780 / 3.70 / $-2.46$  and [Ca/Fe] = 0.4 model from calculations using pure electronic collisions (triangles) and including collisions with \ion{H}{i} according to MGB17 (filled circles), B17u (filled rhombi), 
and MGB17 with no b-b transitions (open rhombi). For comparison, the NLTE corrections were computed with the scaled Drawinian rates (\kH\ = 0.1, open circles). The vertical lines indicate positions of the following lines: 1 = 4425\,\AA, 2 = 5349\,\AA, 3 = 5588\,\AA, 4 = 5590\,\AA, 5 = 5857\,\AA, 6 = 6162\,\AA, 7 = 6169.5\,\AA, 8 = 6439\,\AA, 9 = 6493\,\AA, and 10 = 6499\,\AA.}  \label{Fig:bel_bark}
\end{figure}

Using various collisional recipes, we calculated the NLTE abundance corrections, $\Delta_{\rm NLTE} = \eps{NLTE}-\eps{LTE}$, for the selected lines of \ion{Ca}{i} in the 5780/3.70/$-2.46$ model (Fig.\,\ref{Fig:bel_bark}). They are mostly positive and larger for pure electronic collisions than in any line-formation scenario that includes collisions with \ion{H}{i},  by up to 0.2~dex. In case of MGB17 and B17u, $\Delta_{\rm NLTE}$ does not exceed 0.11 and 0.17~dex, respectively, and the difference in $\Delta_{\rm NLTE}$ between the two options does not exceed 0.06~dex.

We note the particular case of \ion{Ca}{i} 6439\,\AA, for which $\Delta_{\rm NLTE}$ is close to zero in case of MGB17, B17u, and MGB17(no b-b) and even slightly negative in case of scaled Drawinian rates. Here, the same mechanisms are at work as first discussed by \citet{1991MNRAS.251..369D} for the [Fe/H] $\ge -1$ model atmospheres. The line wings are formed in deep layers where over-ionisation
depopulates all \ion{Ca}{i} levels, but the core of \ion{Ca}{i} 6439\,\AA\ in the 5780/3.70/$-2.46$ model is formed at the depths, $\log\tau_{5000} \simeq -0.7$, where the upper level, \eu{4p}{3}{F}{\circ}{}, of the transition is underpopulated to a greater extent than is the lower level, \eu{3d}{3}{D}{}{}, because of photon losses in the line wings. The line source function drops below the Planck function at these depths, resulting in an enhanced
absorption in the line core. The combined effect on the line strength is
that the NLTE abundance correction is small. Sections\,\ref{Sect:stars1} and \ref{Sect:ump} deal with similar phenomena in the formation of stellar \ion{Ca}{i} lines, in particular, the resonance line.




\section{Stellar sample, observations, and atmospheric parameters}\label{Sect:basics}

\begin{table*}  
\caption{Atmospheric parameters and obtained calcium NLTE and LTE abundances of the sample stars.}
\label{startab}
\begin{tabular}{rccrcrrrcrrr}
\noalign{\smallskip} \hline \noalign{\smallskip}
 HD/BD & $\Teff$ & log~g & [Fe/H] & $\Vmic$ & & \multicolumn{2}{c}{\ion{Ca}{i}} & & \multicolumn{3}{c}{\ion{Ca}{ii}} \\
\cline{7-8}
\cline{10-12} \noalign{\smallskip}
       & [K]  &      &         &  [\kms]  & & \multicolumn{1}{c}{[Ca/H]$_{\rm I}$} & $\sigma$~~ ($N$)  & & 8498$^1$ & \multicolumn{1}{c}{[Ca/H]$_{\rm II}^2$} & $\sigma$~~ ($N$) \\
\noalign{\smallskip} \hline \noalign{\smallskip}
 Sun   & 5777 & 4.44 & 0.00    & 0.9 & NLTE &     6.33 & 0.06 (22) &  &    6.27 & 6.40 & 0.05 (7) \\
       &      &      &         &     &  LTE &     6.40 & 0.05 (22) &  &    6.27 & 6.53 & 0.10 (7) \\
 61421 & 6600 & 3.99 & 0.02    & 1.8 & NLTE &  $-$0.07 & 0.05 (20) &  &         & $-$0.01 & 0.05 (3) \\
(Procyon) &      &      &         &     &  LTE &     0.01 & 0.13 (20) & &       &    0.04 & 0.08 (3) \\
 84937 & 6350 & 4.09 & $-$2.12 & 1.7 & NLTE &  $-$1.77 & 0.05 (12) &  & $-$1.77 &  $-$1.60 & 0.03 (2) \\
       &      &      &         &     &  LTE &  $-$1.91 & 0.05 (12) &  & $-$1.53 &  $-$1.80 & 0.04 (2) \\
103095 & 5130 & 4.66 & $-$1.26 & 0.9 & NLTE &  $-$0.98 & 0.04 (21) &  & $-$0.90 & 	 &	  \\
       &      &      &         &     &  LTE &  $-$0.99 & 0.06 (21) &  & $-$0.88 & 	 &        \\
140283 & 5780 & 3.70 & $-$2.46 & 1.6 & NLTE &  $-$2.18 & 0.06 (15) &  & $-$2.12 &  $-$1.96 & 0.04 (2) \\ 
       &      &      &         &     &  LTE &  $-$2.33 & 0.07 (15) &  & $-$1.95 &  $-$2.14 & 0.06 (2) \\ 
$+09^\circ$0352 & 6150 & 4.25& $-$2.09 & 1.3 & NLTE &  $-$1.66 & 0.04 (11) &  & $-$1.59 &  $-$1.63 & 0.01 (2) \\
                &      &     &         &     &  LTE &  $-$1.78 & 0.06 (11) &  & $-$1.46 &  $-$1.83 & 0.01 (2) \\
$-13^\circ$3442 & 6400 & 3.95& $-$2.62 & 1.4 & NLTE &  $-$2.07 & 0.04 ~(6) &  & $-$2.04 &  $-$2.07 &    (1) \\
                &      &     &         &     &  LTE &  $-$2.23 & 0.08 ~(6) &  & $-$1.77 &  $-$2.28 &    (1) \\
\noalign{\smallskip} \hline \noalign{\smallskip}
\multicolumn{12}{l}{Notes. $^1$ [Ca/H] from \ion{Ca}{ii} 8498\,\AA, \ $^2$ from \ion{Ca}{ii} high-excitation lines.} \\
\end{tabular}
\end{table*}

From the sample studied in Paper~I, we selected HD\,61421 (Procyon) and three metal-poor stars, HD~84937, HD~103095, and HD~140283, for which high-resolution observed spectra are available not only in the visible, but also in the IR regions. For these stars, we used the same observational material that was taken from \citet{2003A&A...407..691K} and obtained with the fibre-fed {\'e}chelle spectrograph
FOCES \citep{1998A&AS..130..381P} at the 2.2 m telescope of the Calar Alto observatory, with
a spectral resolution of R = 65\,000 (HD~140283 was observed at R = 40\,000). For these data, the wavelength coverage is 4200\,--\,9000\,\AA. High-quality observations for \ion{Ca}{ii} 8498\,\AA\ in HD\,140283 were taken from the ESO UVESPOP survey \citep{2003Msngr.114...10B}.

To this sample we added BD~$+09^\circ$0352 and BD~$-13^\circ$3442. Their spectra are taken from the CFHT/ESPaDOnS archive\footnote{http://www.cadc-ccda.hia-iha.nrc-cnrc.gc.ca/en/search/} (IDs 1424029 and 1515097-98, respectively). For these data, R = 68\,000 and the wavelength coverage is 3715\,--\,9300\,\AA. 

For the five MP stars, we used atmospheric parameters determined in our earlier study \citep{2015ApJ...808..148S}. In short, a combination of the photometric and spectroscopic methods was applied to derive effective temperatures and surface gravities. The spectroscopic analyses took advantage of the NLTE line-formation modelling for \ion{Fe}{i-ii}, using the method described by \citet{mash_fe}. The iron abundance and microturbulence velocity, $\xi_t$, were determined simultaneously with log~g. 

Atmospheric parameters of Procyon are revised compared with those in Paper~I. We adopted $\Teff$ = 6600~K from \citet{Boyajian2013}, which is based on the bolometric flux measurements and the interferometric angular diameter improved by \citet{2012A&A...540A...5C}. To compute log~g, we used the dynamical mass of 1.478$\pm$0.012 solar mass as derived by \citet{2015ApJ...813..106B} and the linear radius of 2.0362$\pm$0.0145 solar radius from \citet{2012A&A...540A...5C}. The iron abundance and microturbulence velocity were determined from the NLTE analysis of \ion{Fe}{i} and \ion{Fe}{ii} lines. The line list together with accurate line atomic data were compiled by \citet{2015ApJ...808..148S}. Excluding the strong lines with an observed equivalent width of $W_{obs} >$ 120~m\AA, we have 31 lines of \ion{Fe}{i} and 15 lines lines of \ion{Fe}{ii}. In a line-by-line differential analysis relative to the Sun, $\xi_t$ = 1.8\,\kms\ appears to provide the faintest slope of the \ion{Fe}{i}-based abundance versus line strength trend. This value coincides with the earlier determination of \citet{2003A&A...407..691K}, which was also used in Paper~I. The iron differential abundances based on lines of \ion{Fe}{i} and \ion{Fe}{ii}, [Fe/H]$_{\rm I} = -0.01\pm0.10$ and [Fe/H]$_{\rm II} = 0.05\pm0.04$, agree within 0.06~dex. Their mean is adopted as a final iron abundance.

Four of our sample stars were used by  \citet{2015A&A...582A..49H} as benchmark stars. For two of them, HD~84937 and Procyon, the adopted atmospheric parameters agree well in the two studies. For the other two stars, HD~103095 and HD~140283, their interferometric effective temperatures are not recommended by  \citet{2015A&A...582A..49H} to "be used as a reference for calibration or validation purposes".

Stellar atmosphere parameters are given in Table\,\ref{startab}.



\section{\ion{Ca}{i} versus \ion{Ca}{ii} in the sample stars} \label{Sect:stars1}

\begin{table*}
\caption{Atomic data for the selected \ion{Ca}{i} and \ion{Ca}{ii} lines and  
 the NLTE and LTE abundances, $\eps{\odot}$, determined from the line profiles
in the Kitt Peak Solar Atlas \citep{Atlas}. } \label{line_list} 
\begin{center}
\begin{tabular}{lrcrlclrrrr}
\hline\noalign{\smallskip}
 \multicolumn{1}{c}{$\lambda$} &    mult & E$_{low}$ &  \multicolumn{1}{c}{$\log gf$} & Ref &\multicolumn{1}{c}{log C$_6$} & Ref & \multicolumn{2}{c}{$\eps{\odot}$} & $\Delta_{\rm NLTE}$
& W$_{\odot}$ \\
\cline{8-9}
 \multicolumn{1}{c}{[\AA]} & & [eV] &  & & & & NLTE & LTE & [dex] & [m\AA] \\
\hline
\multicolumn{1}{c}{1} & \multicolumn{1}{c}{2} & 3 & \multicolumn{1}{c}{4} & \multicolumn{1}{c}{5} & \multicolumn{1}{c}{6}
 & \multicolumn{1}{c}{7} & \multicolumn{1}{c}{8} & 9 & \multicolumn{1}{c}{10} & \multicolumn{1}{c}{11} \\
\hline \noalign{\smallskip}
\multicolumn{2}{c}{\ion{Ca}{i}} &  & & & & & & & & \\
  4226.73  &   2 &0.00& ~0.244   & SG66  & $-$31.23 & $ABO$ & 6.33   &  6.30    & 0.03   &  $>$1\,\AA \\
  4425.44  &   4 &1.87& $-$0.358 & SON75 & $-$30.23 & $ABO$ & 6.23   &  6.23    & 0.00   &    174 \\
  4578.56  &  23 &2.51& $-$0.697 & SR81  & $-$30.30 & S81 &  6.33  &  6.37    & $-$0.04   &     87\\
  5261.71  &  22 &2.51& $-$0.579 & SR81  & $-$30.86 & S81 &  6.36  &  6.41    & $-$0.05   &    102 \\
  5349.47  &  33 &2.70& $-$0.310 & SR81  & $-$31.45 & S81 &  6.32  &  6.43    & $-$0.11   &     97 \\
  5512.98$^1$&   &2.92& $-$0.464 & S88   & $-$30.61 & S81 &  6.38  &  6.41    & $-$0.03   &     89 \\
  5588.76  &  21 &2.51& ~0.358   & SR81  & $-$31.39 & S81 &  6.27  &  6.39    & $-$0.12   &    171 \\
  5590.12  &  21 &2.51& $-$0.571 & SR81  & $-$31.39 & S81 &  6.33  &  6.42    & $-$0.09   &     97 \\
  5857.45  &  47 &2.92&   ~0.240 & SR81  & $-$30.61 & S81 &  6.30  &  6.41    & $-$0.11   &    160 \\
  5867.57$^2$&   &2.92& $-$1.570 & S88   & $-$30.97 &VALD &  6.39  &  6.37    &    0.02   &     26 \\
  6122.22  &   3 &1.88& $-$0.315 & SON75 & $-$30.30 & $ABO$ &  6.36  &  6.39    & $-$0.03   &    228 \\
  6162.17  &   3 &1.89& $-$0.089 & SON75 & $-$30.30 & $ABO$ &  6.38  &  6.38    &    0.00   &    289 \\
  6161.29  &  20 &2.51& $-$1.266 & SR81  & $-$30.48 & S81 &  6.40  &  6.39    &    0.01   &     67 \\
  6166.44  &  20 &2.51& $-$1.142 & SR81  & $-$30.48 & S81 &  6.41  &  6.41    &    0.00   &     77 \\
  6169.06  &  20 &2.51& $-$0.797 & SR81  & $-$30.48 & S81 &  6.41  &  6.45    & $-$0.04   &    100 \\
  6169.56  &  20 &2.51& $-$0.478 & SR81  & $-$30.48 & S81 &  6.37  &  6.41    & $-$0.04   &    129 \\
  6439.07  &  18 &2.51&   ~0.390 & SR81  & $-$31.58 & S81 &  6.25  &  6.38    & $-$0.13   &    188 \\
  6471.66  &  18 &2.51& $-$0.686 & SR81  & $-$31.58 & S81 &  6.32  &  6.44    & $-$0.12  &     97 \\
  6493.78  &  18 &2.51& $-$0.109 & SR81  & $-$31.58 & S81 &  6.27  &  6.41    & $-$0.14   &    139\\
  6499.65  &  18 &2.51& $-$0.818 & SR81  & $-$31.58 & S81 &  6.35  &  6.47    & $-$0.12  &     89 \\
  6449.81  &  19 &2.51& $-$0.502 & SR81  & $-$31.45 & S81 &  6.23  &  6.45    & $-$0.22  &    106 \\
  6455.60  &  19 &2.51& $-$1.34  & S88   & $-$31.45 & S81 &  6.32  &  6.39    & $-$0.07  &     59 \\
  6572.78  &   1 &0.00& $-$4.24  & DIK97 & $-$31.54 & $ABO$ &  6.27  &  6.24    &    0.03   &     31 \\
\multicolumn{2}{c}{\ion{Ca}{ii}} &  & & & & & & & & \\
  3706.02  &   3 &3.12& $-$0.40  & VALD  & $-$31.34 & VALD&        &          &           &  \\
  3933.66  &   1 &0.00&  ~0.105  & T89   & $-$31.94 & $ABO$ &        &          &           & $>$1\,\AA \\
  5339.19  &  20 &8.40& $-$0.079 & OP    & $-$30.31 & VALD&  6.38  &   6.38   &    0.00   &      6 \\
  6456.91  &  19 &8.40&   ~0.412 & OP    & $-$30.77 & VALD&  6.41  &  6.42    & $-$0.01   &     16\\
  8248.80  &  13 &7.48&   ~0.556 & OP    & $-$30.85 & Sun &  6.42  &  6.58    & $-$0.16   &     65\\
  8254.70  &  13 &7.48& $-$0.398 & OP    & $-$30.85 & Sun &  6.45  &  6.49    & $-$0.04   &     18  \\
  8498.02  &   2 &1.69& $-$1.416 & T89   & $-$31.51 & $ABO$ & 6.27   &  6.27    &    0.00   & $>$1\,\AA \\
  8542.31  &   2 &1.70& $-$0.463 & T89   & $-$31.51 & $ABO$ & 6.27   &  6.27    &    0.00   & $>$1\,\AA \\
  8662.30  &   2 &1.69& $-$0.723 & T89   & $-$31.51 & $ABO$ & 6.25   &  6.25    &    0.00   & $>$1\,\AA \\
  8912.07$^3$&   &7.05&   ~0.636 & OP    & $-$31.10 & Sun &  6.35  &  6.63    & $-$0.28   &    111\\
  8927.36$^4$&   &7.05&   ~0.811 & OP    & $-$31.10 & Sun &  6.31  &  6.62    & $-$0.31   &    121\\
  9890.70$^5$&   &8.40&   ~1.270 & OP    & $-$31.34 &VALD &  6.43  &  6.51    & $-$0.08   &     70 \\
\noalign{\smallskip}\hline \noalign{\smallskip}
\multicolumn{11}{l}{Column 2 contains the multiplet numbers accordingly to \citet{1972mtai.book.....M}.} \\
\multicolumn{11}{l}{Column 11 contains solar equivalent widths.} \\
\multicolumn{3}{l}{$^1$ \eu{4p}{1}{P}{\circ}{} - \eu{4p^2}{1}{S}{}{}} & \multicolumn{4}{l}{SG66 = \citet{1966PhRv..145...26S} } & \multicolumn{4}{l}{T89 = \citet{1989PhRvA..39.4880T} } \\
\multicolumn{3}{l}{$^2$ \eu{4p}{1}{P}{\circ}{} - \eu{6s}{1}{S}{}{}}  & \multicolumn{4}{l}{SON75 = \citet{1975A&A....38....1S}} & \multicolumn{4}{l}{OP = \citet{1994MNRAS.266..805S} } \\
\multicolumn{3}{l}{$^3$ \eu{4d}{2}{D}{}{3/2} - \eu{4f}{2}{F}{\circ}{5/2}}  & \multicolumn{4}{l}{SR81 =  \citet{1981JPhB...14.4015S}} & \multicolumn{4}{l}{S81 = \citet{1981A&A...103..351S} } \\
\multicolumn{3}{l}{$^4$ \eu{4d}{2}{D}{}{5/2}  - \eu{4f}{2}{F}{\circ}{5/2,7/2}} & \multicolumn{4}{l}{S88   = \citet{1988JPhB...21.2827S}} & \multicolumn{4}{l}{Sun = solar line profile fitting} \\
\multicolumn{3}{l}{$^5$ \eu{4f}{2}{F}{\circ}{}  - \eu{5g}{2}{G}{}{}} & \multicolumn{4}{l}{DIK97 = \citet{1997ZPhyD..41..125D}} & \multicolumn{4}{l}{} \\
\end{tabular}
\end{center}
\end{table*}

In this section, we derive abundances from lines of \ion{Ca}{i} and \ion{Ca}{ii} in the Sun and selected stars and inspect the abundance differences between the two ionisation stages. For the stars, we applied a line-by-line differential NLTE and
LTE approach, in the sense that stellar line abundances were compared with individual abundances of their solar counterparts. 
The NLTE abundances were calculated using rate coefficients for \ion{Ca}{i} + \ion{H}{i} collisions from \citet{ca1_hydrogen}.

\subsection{Line list, line atomic data, and the codes}

We employed the line list and the line atomic data that were comprehensively tested in Paper~I. In addition, Table\,\ref{line_list} includes lines of \ion{Ca}{ii}, which are used in Sect.\,\ref{Sect:ump} to derive Ca abundances of the UMP stars.
Accurate laboratory $gf$-values are available for all investigated lines of \ion{Ca}{i} and the low-excitation lines of \ion{Ca}{ii},  with \Eexc $\leq$ 3.12~eV; the sources of data are indicated in Table\,\ref{line_list}. For high-excitation lines of \ion{Ca}{ii},  with \Eexc $\geq$ 7.05~eV, $gf$-values based on OP calculations were adopted. 
For the \ion{Ca}{ii} IR triplet lines we account for the isotope structure with the isotope shifts measured by \citet{1998EPJD....2...33N}. For the Ca isotope abundance ratios the solar system values from \citet{Lodders2009} were adopted.
Isotope shifts are much smaller ($<$10\,m\AA) for the resonance lines in
\ion{Ca}{i} and \ion{Ca}{ii} \citep{2004PhRvA..69a2711L} and
we treat them as single lines. No data are available for the
remaining Ca lines. We neglected the hyperfine structure of Ca lines because
the fractional abundance of the only odd isotope \iso{43}{Ca} is very low (0.135\%).
For 17 \ion{Ca}{i} lines, the van der Waals $C_6$ values were computed from damping parameters given by \citet{1981A&A...103..351S} and based on the measured parameters for broadening by helium. For the remaining
lines, we adopted $C_6$ values based on the theory of collisional broadening by atomic hydrogen treated by \citet{1995MNRAS.276..859A,1997MNRAS.290..102B,1998MNRAS.300..863B}, and \citet{1998MNRAS.296.1057B}. Hereafter, these four important papers are referred to collectively as $ABO$. When the line is not available in any cited source, we rely on Kurucz's\footnote{\tt http://kurucz.harvard.edu/atoms.html} $C_6$ values, which are accessible in the VALD database \citep{2015PhyS...90e4005R}.

All our results are based on line-profile analysis. The theoretical spectra were computed with the code {\sc synthV-NLTE} \citep{Ryabchikova2015}, which uses the departure coefficients from the
DETAIL program \citep{detail}. In turn, {\sc synthV-NLTE} was integrated within the
{\sc idl binmag}3 code\footnote{\tt http://www.astro.uu.se/$\sim$oleg/download.html} written by O. Kochukhov, finally allowing us to determine the best fit to the observed line profiles.

\subsection{Analysis of solar \ion{Ca}{i} and \ion{Ca}{ii} lines}\label{Sect:sun}

\begin{figure}
\resizebox{88mm}{!}{\includegraphics{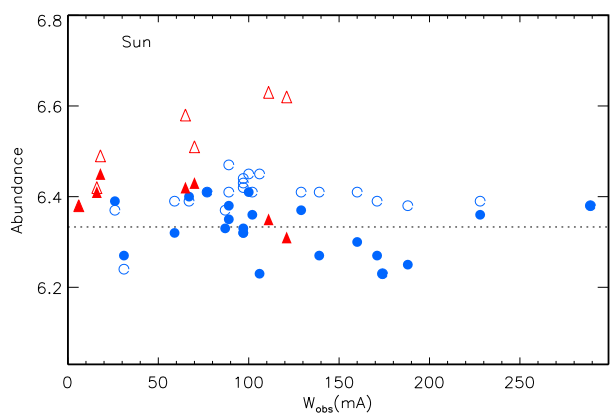}}

\caption[]{Solar NLTE (filled symbols) and LTE (open symbols) calcium abundances, $\eps{}$, from the \ion{Ca}{i} (circles) and \ion{Ca}{ii} (triangles) lines plotted as a function of the line equivalent width. The mean NLTE abundance derived from the \ion{Ca}{i} lines is shown by the dotted line.}  \label{Fig:ca_solar}
\end{figure}

We used solar flux observations taken from the Kitt Peak Solar Atlas \citep{Atlas}. The calculations were performed with the MARCS solar model atmosphere ($\Teff$ = 5777~K, log~g = 4.44, [Fe/H] = 0) and a depth-independent microturbulence of 0.9\,\kms. When applying {\sc synthV-NLTE}+{\sc binmag}3, we fixed a spectral resolution at R = 500\,000 and a rotation velocity at $V \sin i$ = 1.8\,\kms. Free parameters in line-profile fitting are the Ca abundance and the radial-tangential macroturbulence. 

The LTE and NLTE abundances determined from 22 subordinate lines of \ion{Ca}{i} and seven high-excitation lines of \ion{Ca}{ii} are presented in Table\,\ref{line_list} and Fig.\,\ref{Fig:ca_solar}. 
As discussed in Paper~I, NLTE leads to strengthening most of \ion{Ca}{i} lines in the solar atmosphere. The mechanisms are similar to those for \ion{Ca}{i} 6439\,\AA\ in the 5780/3.70/$-2.46$ model (Sect.\,\ref{Sect:effect}), that is,
dropping the line source function below the Planck function at the line core formation depths results in the enhanced absorption in the line core, and this prevails over weakening the line wings formed in deep layers where over-ionisation depopulates all \ion{Ca}{i} levels. The NLTE effects appear to be stronger when quantum-mechanical rate coefficients are applied for \ion{Ca}{i} + \ion{H}{i} collisions compared with the effect for the scaled (\kH\ = 0.1) Drawinian rates. For example, the corresponding NLTE corrections amount to $\Delta_{\rm NLTE} = -0.07$ and $-0.03$~dex for a moderate-strength line \ion{Ca}{i} 6455\,\AA\ ($W_{obs}$ = 59\,m\AA). 

The \ion{Ca}{i}-based NLTE abundance obtained in this study, $\eps{CaI}$ = 6.33, is lower than the values calculated in Paper~I with \kH\ = 0 (no \ion{Ca}{i} + \ion{H}{i} collisions) and \kH\ = 0.1, by 0.04~dex and 0.03~dex, respectively. This is mostly due to the different treatment of \ion{Ca}{i} + \ion{H}{i} collisions, but not using different solar model atmospheres, that is, to the MARCS model in this study and the MAFAGS \citep{1997A&A...323..909F} model in Paper~I and to not neglecting \ion{Ca}{i} 7326\,\AA\ in this study. Hereafter, the abundances are placed on a classical scale with $\eps{H} = 12$. 

For \ion{Ca}{ii}, NLTE leads to fairly consistent abundances from seven high-excitation lines (Fig.\,\ref{Fig:ca_solar}), in contrast to LTE, where $\sigma$ = 0.10~dex is twice larger. Hereafter, the sample standard deviation, $\sigma = \sqrt{\Sigma(\overline{x}-x_i)^2 / (N-1)}$, determines 
the dispersion in the single line measurements around the mean, where $N$ is the number of measured lines. We recall that \ion{Ca}{ii} + \ion{H}{i} collisions are treated in this study using the scaled (\kH\ = 0.1) Drawinian rates. Therefore, the obtained mean abundance, $\eps{CaII}$ = 6.40, coincides with the corresponding value in Paper~I. It is worth noting that a lower abundance, by 0.13~dex, is determined from the \ion{Ca}{ii} 8498\,\AA\ line wings that are free of the NLTE effects. An abundance discrepancy between the \ion{Ca}{ii} high-excitation lines and \ion{Ca}{ii} 8498\,\AA\ can be caused by the uncertainty in predicted $gf$-values for the first group of lines or/and the uncertainty in predicted van der Waals damping constant for \ion{Ca}{ii} 8498\,\AA.

The obtained \ion{Ca}{i}-based NLTE abundance, $\eps{CaI}$ = 6.33, agrees well with the meteoritic value, $\eps{met}$ = 6.31$\pm$0.02 \citep{Lodders2009}.
Our solar results can also be compared with calculations of \citet{2015A&A...573A..25S}.
Using different line lists (11 lines of \ion{Ca}{i} and 5 lines of \ion{Ca}{ii}) compared with ours and the same solar MARCS model, they obtained the NLTE abundances $\eps{CaI}$ = 6.26 and $\eps{CaII}$ = 6.28. Their recommended solar Ca abundance, $\eps{Ca}$ = 6.32$\pm$0.03, is based on the LTE abundances calculated with the 3D hydrodynamic model of the solar photosphere, to which the NLTE corrections from the 1D model were applied.


\subsection{Stellar abundances from lines of \ion{Ca}{i} and \ion{Ca}{ii}}\label{Sect:vmp}

\begin{figure}
\resizebox{88mm}{!}{\includegraphics{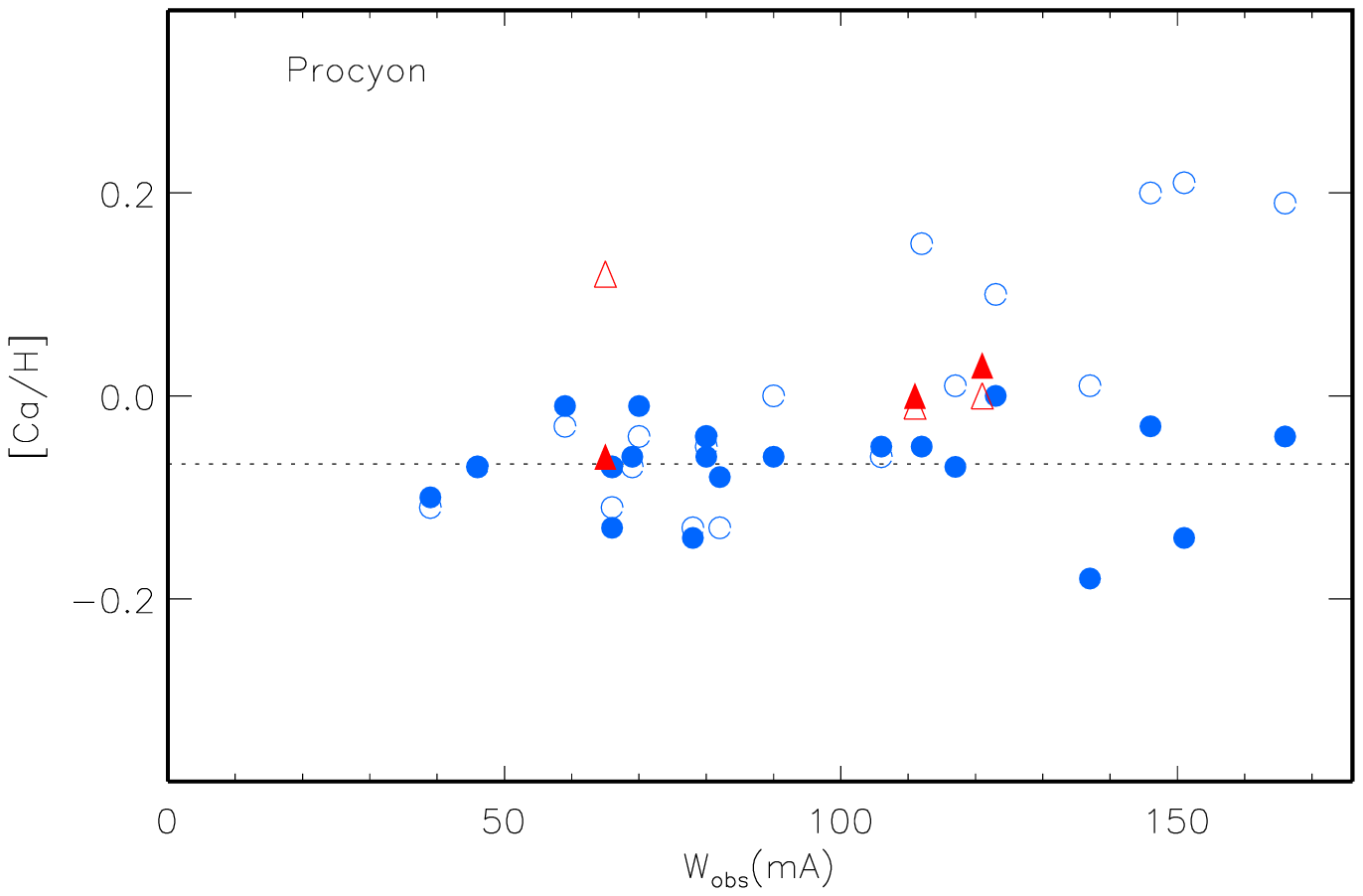}}

\caption[]{NLTE (filled symbols) and LTE (open symbols) [Ca/H] abundances determined from the \ion{Ca}{i} (circles) and \ion{Ca}{ii} (triangles) lines in Procyon. Note the steep trend of the \ion{Ca}{i}-based LTE values with line strength above $W_{obs}$ = 100\,m\AA. The dotted line indicates the mean [Ca/Fe]$_{\rm I}$ NLTE value.} \label{Fig:procyon}
\end{figure}

When we applied {\sc synthV-NLTE}+{\sc binmag}3, we fixed a spectral resolution for a given star at the value indicated in Sect.\,\ref{Sect:basics}. Only for Procyon, rotational broadening, with $V \sin i = 2.6$\,\kms\ from \citet{Fuhrmann1998}, and broadening 
by macroturbulence were treated separately. The remaining stars are slow rotators, $V \sin i
\precsim 1.5$\,\kms, and therefore the rotational velocity and macroturbulence value cannot be separated at the spectral
resolving power of our spectra. We therefore consider their overall effect as
radial-tangential macroturbulence, and together with the Ca abundance, they are free parameters in line-profile fitting.  

For each star, we determined abundances from the subordinate lines of \ion{Ca}{i}, the high-excitation lines of \ion{Ca}{ii}, where available, and for the MP stars, from \ion{Ca}{ii} 8498\,\AA. Table\,\ref{startab} presents the mean NLTE and LTE abundances. For \ion{Ca}{i}, the difference in the mean abundance between the two NLTE scenarios exceeds 0.02~dex at no point. The NLTE data in Table\,\ref{startab} correspond to the MGB17 scenario.  


We first focus on Procyon, for which our earlier study \citep{mash_ca} obtained an abundance difference of more than 0.2~dex between the two ionisation stages, \ion{Ca}{i} and \ion{Ca}{ii}, independently of the line-formation scenario, and the determined \ion{Ca}{i}-based abundance was clearly subsolar, although a success was achieved in removing a steep trend with line strength among strong \ion{Ca}{i} lines seen in LTE. In this study, we applied a 90~K higher effective temperature, and this appears the key to the solution of the abundance problems. The NLTE abundances from \ion{Ca}{i} and \ion{Ca}{ii} agree within 0.06~dex (Table\,\ref{startab}), and [Ca/H]$_{\rm I} = -0.07$ ($\sigma = 0.05$) is close to the solar value. 
As in Paper~I,  we obtained consistent NLTE abundances from  \ion{Ca}{i} lines of different strength, in contrast to the LTE abundances, which reveal a steep increasing trend at $W_{obs} >$ 110\,m\AA\ (Fig.\,\ref{Fig:procyon}). 

NLTE largely removes the line-to-line scatter for \ion{Ca}{ii}.

\begin{figure}
\resizebox{88mm}{!}{\includegraphics{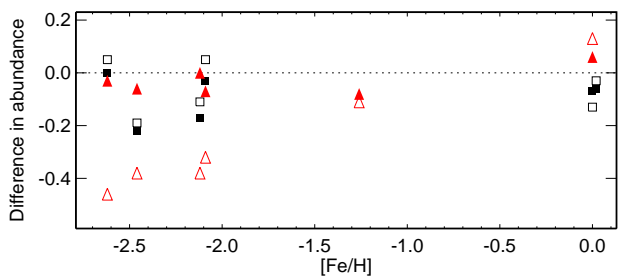}}

\caption[]{NLTE (filled symbols) and LTE (open symbols) differences between [Ca/H]$_{\rm I}$ and [Ca/H]$_{\rm II}$ for the sample stars. The squares and triangles correspond to the high-excitation lines of \ion{Ca}{ii} and \ion{Ca}{ii} 8498\,\AA, respectively.} \label{Fig:stars}
\end{figure}

NLTE also works well for our MP stars. For every star, it reduces the line-to-line scatter for both \ion{Ca}{i} and \ion{Ca}{ii} compared with LTE. NLTE leads to higher abundances from lines of \ion{Ca}{i} and lower abundance from \ion{Ca}{ii} 8498\,\AA, such that these two abundance indicators provide consistent results, in contrast to LTE, where the abundance difference (\ion{Ca}{i} -- \ion{Ca}{ii} 8498\,\AA) grows towards lower metallicity and reaches $-0.46$~dex in BD~$-13^\circ$3442 ([Fe/H] = $-2.62$, Fig.\,\ref{Fig:stars}). 

For \ion{Ca}{ii} 8912, 8927\,\AA\ in the VMP stars, NLTE leads to strengthened lines and negative NLTE abundance corrections, which are smaller in absolute value than those for the solar lines, resulting in positive abundance differences, [Ca/H]$_{\rm NLTE}$ -- [Ca/H]$_{\rm LTE}$.
Their magnitudes are as large as the magnitude for the \ion{Ca}{i} lines. Therefore, the NLTE and LTE abundance differences [Ca/H]$_{\rm I}$ -- [Ca/H]$_{\rm II}$ are close together for each star (Fig.\,\ref{Fig:stars}). They are within 1$\sigma$ for BD~$+09^\circ$0352 and BD~$-13^\circ$3442, but amount to $-0.17$ and $-0.22$~dex for HD~84937 and HD~140283, respectively. 

To determine whether the discrepancy in abundance between different lines of \ion{Ca}{ii} might be caused by an uncertainty in log~g, we calculated NLTE abundances of HD~84937 using an 0.1~dex lower surface gravity. A decrease of log~g leads to stronger \ion{Ca}{ii} lines and lower derived abundances. For \ion{Ca}{ii} 8912, 8927\,\AA, the abundance shifts amount to $-0.04$~dex and $-0.03$~dex, respectively. The NLTE abundance from \ion{Ca}{ii} 8498\,\AA\ is the same regardless of whether we use log~g = 4.09 or 3.99. A change of 0.1~dex in log~g for HD~84937 leads to a minor change of 0.04~dex in the NLTE abundance difference between \ion{Ca}{ii} lines and does not help to achieve consistent abundances from different lines of \ion{Ca}{ii}.

\section{\ion{Ca}{i} versus \ion{Ca}{ii} in the UMP stars} \label{Sect:ump}

\begin{table*}  
\caption{Atmospheric parameters and obtained calcium NLTE and LTE abundances of the UMP stars.}
\label{star_UMP}
\begin{tabular}{ccclccclccl}
\noalign{\smallskip} \hline \noalign{\smallskip}
 HE & $\Teff$ & log~g & [Fe/H] & $\Vmic$ & Ref & \multicolumn{2}{c}{NLTE, $\eps{Ca}$} & & \multicolumn{2}{c}{LTE, $\eps{Ca}$} \\
\cline{7-8}
\cline{10-11} \noalign{\smallskip}
    & [K]     &       &        & [\kms]  &     & \multicolumn{1}{c}{\ion{Ca}{i} 4226\,\AA} & \multicolumn{1}{c}{\ion{Ca}{ii}} & & \multicolumn{1}{c}{\ion{Ca}{i} 4226\,\AA} & \multicolumn{1}{c}{\ion{Ca}{ii}} \\
\noalign{\smallskip} \hline \noalign{\smallskip}
0107-5240 & 5100 & 2.2 & $-$5.3  & 2.2 & CGK04 & 1.28 & 1.38           & & 0.94 & 1.41 	   \\
0557-4840 & 4900 & 2.2 & $-$4.75 & 1.8 & NCK07 & 1.94 & 2.20 (5, 0.05) & & 1.78 & 2.55 (5, 0.37) \\
1327-2326 & 6180 & 3.7 & $-$5.45 & 1.7 & AFC06 & 1.16 & 1.18 (2, 0.02) & & 0.91 & 1.33 (2, 0.06) \\
\noalign{\smallskip} \hline \noalign{\smallskip}
\multicolumn{11}{l}{Ref: AFC06 = \citet{Aoki_he1327}, CGK04 = \citet{HE0107_ApJ}, NCK07 = \citet{Norrisetal:2007}} \\
\multicolumn{11}{l}{Numbers in parenthesis indicate the number of lines measured and $\sigma$.} \\
\end{tabular}
\end{table*}

In this section, we determine Ca absolute abundances of the three UMP stars listed in Table\,\ref{star_UMP}. Five lines of \ion{Ca}{ii}, that is, the resonance line at 3933\,\AA, the IR triplet lines, and the low-excitation line at 3706\,\AA, were observed in HE\,0557-4840 and two lines, \ion{Ca}{ii} 3933 and 8498\,\AA, in HE\,1327-2326. We inspected whether different lines of \ion{Ca}{ii} in a given star yield consistent abundances. In all three stars, neutral calcium is represented only by the \ion{Ca}{i} 4226\,\AA\ resonance. For each star, we checked the \ion{Ca}{i}/\ion{Ca}{ii} ionisation equilibrium with given atmospheric parameters.

For HE\,0557-4840,  HE\,0107-5240, and HE\,1327-2326, we use reduced spectra from the VLT/UVES\footnote{http://archive.eso.org/wdb/wdb/adp/phase3-main/query} archive (IDs 380.D-0040(A);  0.D-0009(A), 076.D-0165(A);  and 075.D-0048(A), 077.D-0035(A), respectively) and the earlier study of  \citet{2009ApJ...698..410K}, in which one of us (L.M.) took part. 
Atmospheric parameters were taken from the literature, as indicated in Table\,\ref{star_UMP}.

 \begin{figure}
\resizebox{88mm}{!}{\includegraphics{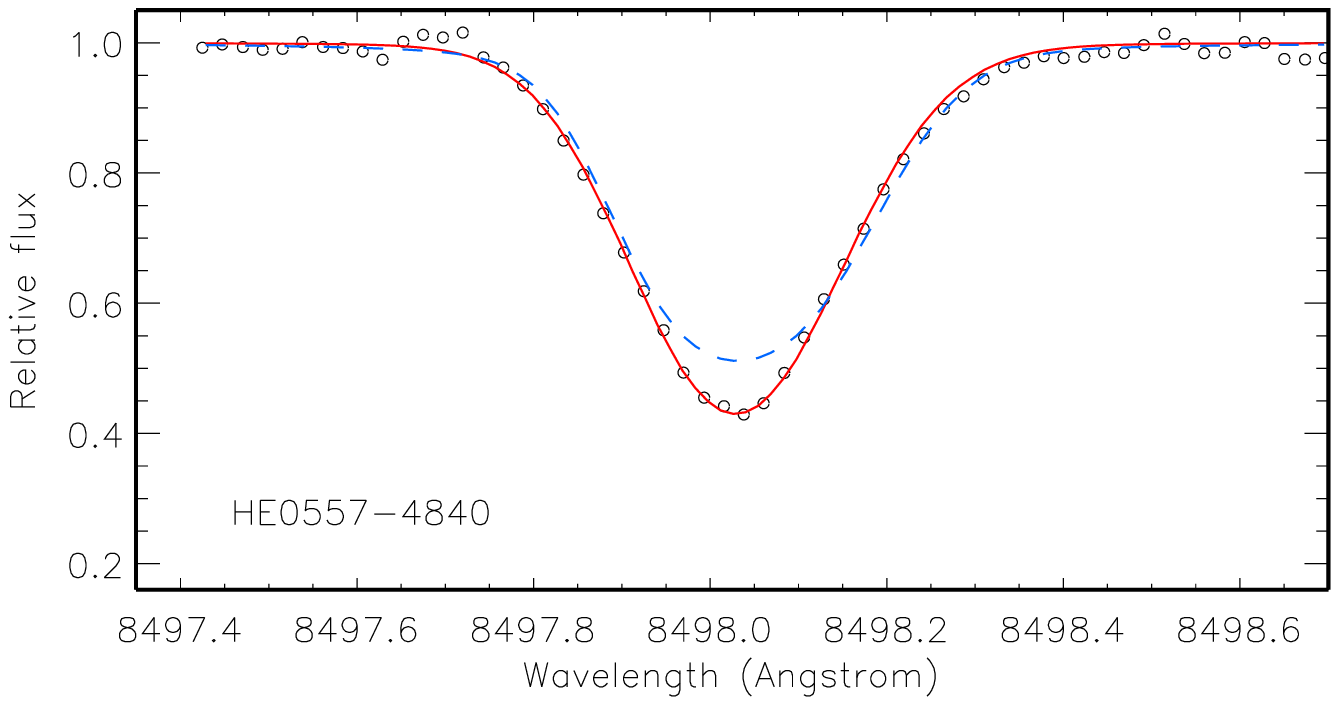}}

\vspace{-3mm}
\resizebox{88mm}{!}{\includegraphics{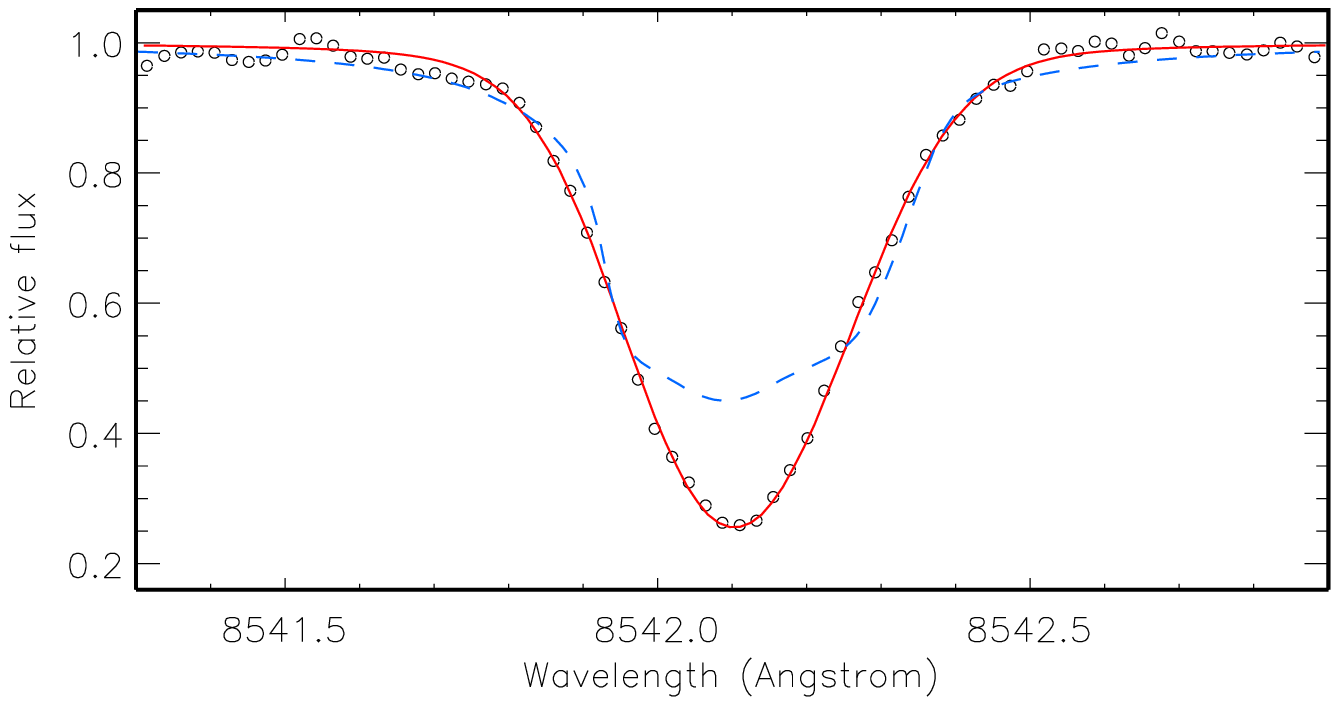}}

\vspace{-3mm}
\resizebox{88mm}{!}{\includegraphics{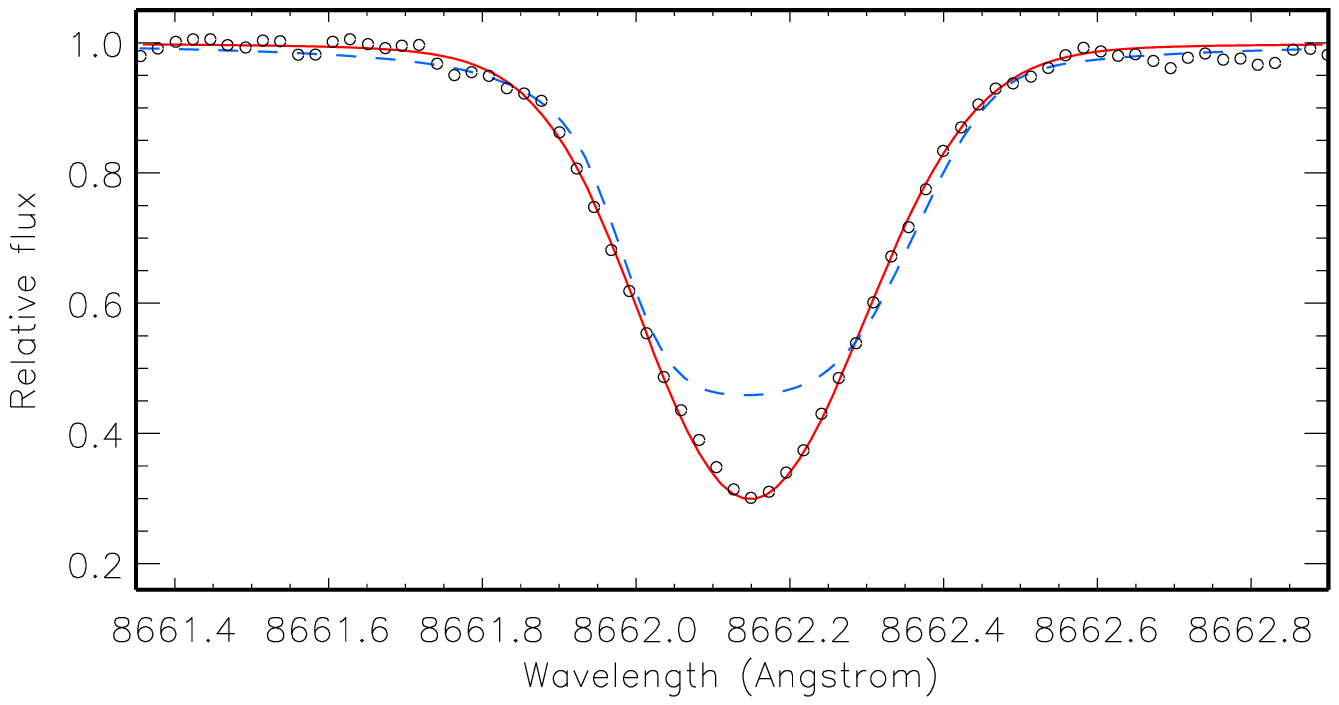}}
\caption[]{Best NLTE (continuous curve) fits of the \ion{Ca}{ii} IR triplet lines in HE\,0557-4840 (open circles). The obtained NLTE abundances are $\eps{Ca}$ = 2.21, 2.13, and 2.15 for \ion{Ca}{ii} 8498, 8542, and 8662\,\AA, respectively. 
The LTE profiles (dashed curve) computed with 0.57~dex higher Ca abundance compared with the corresponding NLTE one. 
 See text for more details.
} \label{Fig:caIR}
\end{figure}

Pronounced departures from LTE are found for the \ion{Ca}{ii} IR triplet lines in HE\,0557-4840 (Fig.\,\ref{Fig:caIR}), such that the observed profile cannot be reproduced in LTE for any of the lines. 
In Fig.\,\ref{Fig:caIR} we show the LTE profiles for the weakest and the strongest of the triplet lines. First, we note the shift in wavelength between the NLTE and LTE profile that is due to different NLTE affecting the different isotopic components of the line. Second, the LTE profile is much shallower than the observed profile, even when no external broadening by macro-turbulence and instrumental broadening is applied, as shown for \ion{Ca}{ii} 8542 and 8662\,\AA.
When using the observed equivalent widths, we obtain an LTE abundance that is substantially higher than the NLTE abundance, for example, by 0.67~dex for \ion{Ca}{ii} 8498\,\AA. In contrast, NLTE leads to weakening \ion{Ca}{ii} 3706\,\AA\ and positive abundance correction of $\Delta_{\rm NLTE}$ = 0.18~dex. For \ion{Ca}{ii} 3933\,\AA, the NLTE effects are minor, with $\Delta_{\rm NLTE} <$ 0.01~dex. We find fairly consistent NLTE abundances from different lines of \ion{Ca}{ii} in HE\,0557-4840, with $\sigma$ = 0.05~dex (Table\,\ref{star_UMP}). The \ion{Ca}{i} resonance line gives a lower NLTE abundance than that deduced from \ion{Ca}{ii} lines, by 0.26~dex. These results differ from  those of \citet{Norrisetal:2007} because (i) using accurate quantum-mechanical data on \ion{Ca}{i} + \ion{H}{i} collisions leads to smaller positive NLTE correction for \ion{Ca}{i} 4226\,\AA\ compared with that for the scaled Drawinian rates, by 0.10~dex in the MGB17 scenario; (ii) we used absolute, but not differential abundances; (iii) in addition to \ion{Ca}{ii} 3933 and 3706\,\AA, we used the \ion{Ca}{ii} IR triplet lines. 

Both lines of \ion{Ca}{ii} and the \ion{Ca}{i} resonance line in HE\,1327-2326 give NLTE abundances that are  consistent within 0.02~dex, in contrast to LTE, where the abundance difference between \ion{Ca}{ii} and \ion{Ca}{i} amounts to 0.42~dex.

For HE\,0107-5240, NLTE largely removes an abundance discrepancy of 0.47~dex between \ion{Ca}{ii} 3933\,\AA\ and \ion{Ca}{i} 4226\,\AA, which was obtained at the LTE assumption. 

\begin{figure}
\resizebox{88mm}{!}{\includegraphics{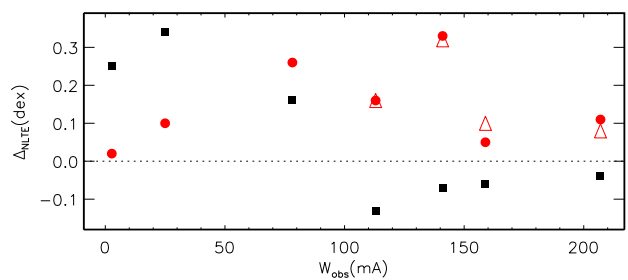}}
\caption[]{
NLTE abundance corrections for \ion{Ca}{i} 4226\,\AA\ (squares) in most of the investigated MP stars as a function of line strength and the corresponding absolute abundance differences between \ion{Ca}{ii} and \ion{Ca}{i} 4226\,\AA\ (circles) and between the \ion{Ca}{i} subordinate and resonance lines (triangles).
}  \label{Fig:ca4226_dnlte}
\end{figure}

In Fig.\,\ref{Fig:ca4226_dnlte} we display an abundance difference between \ion{Ca}{i} 4226\,\AA\ and \ion{Ca}{ii} lines not only for the UMP stars, but also for our four VMP stars. For the latter group, we used the abundances from \ion{Ca}{ii} 8498\,\AA, which agree well with the values deduced from the \ion{Ca}{i} subordinate lines (Sect.\,\ref{Sect:vmp}) and thus are reliable. The \ion{Ca}{i} resonance line yields a  slightly lower abundance than the \ion{Ca}{ii}-based one, by up to 0.1~dex, when it is either weak ($W_{obs}$ = 2.7 and 25\,m\AA\ in HE\,1327-2326 and HE\,0107-5240, respectively) or strong enough ($W_{obs} >$ 150\,m\AA\ in HD~84937 and BD~$+09^\circ$0352). In the other three stars, HD~140283, BD~$-13^\circ$3442, and HE~0557-4840, \ion{Ca}{i} 4226\,\AA\ is of moderate strength and gives substantially lower abundance than that from \ion{Ca}{ii} lines and also from subordinate lines of \ion{Ca}{i}, where available. For example, in HD~140283, the abundance difference between \ion{Ca}{i} resonance and subordinate lines amounts to $-0.32$~dex and $-0.33$~dex between \ion{Ca}{i} 4226\,\AA\ and \ion{Ca}{ii} 8498\,\AA. The problem of the  underestimated abundance from \ion{Ca}{i} 4226\,\AA\ was noted in our Paper~I for the [Fe/H] $\simeq -2$ dwarf stars and highlighted by \citet{2012A&A...541A.143S} for their sample of VMP turnoff stars and giants.

\begin{figure*}
\resizebox{60mm}{!}{\includegraphics{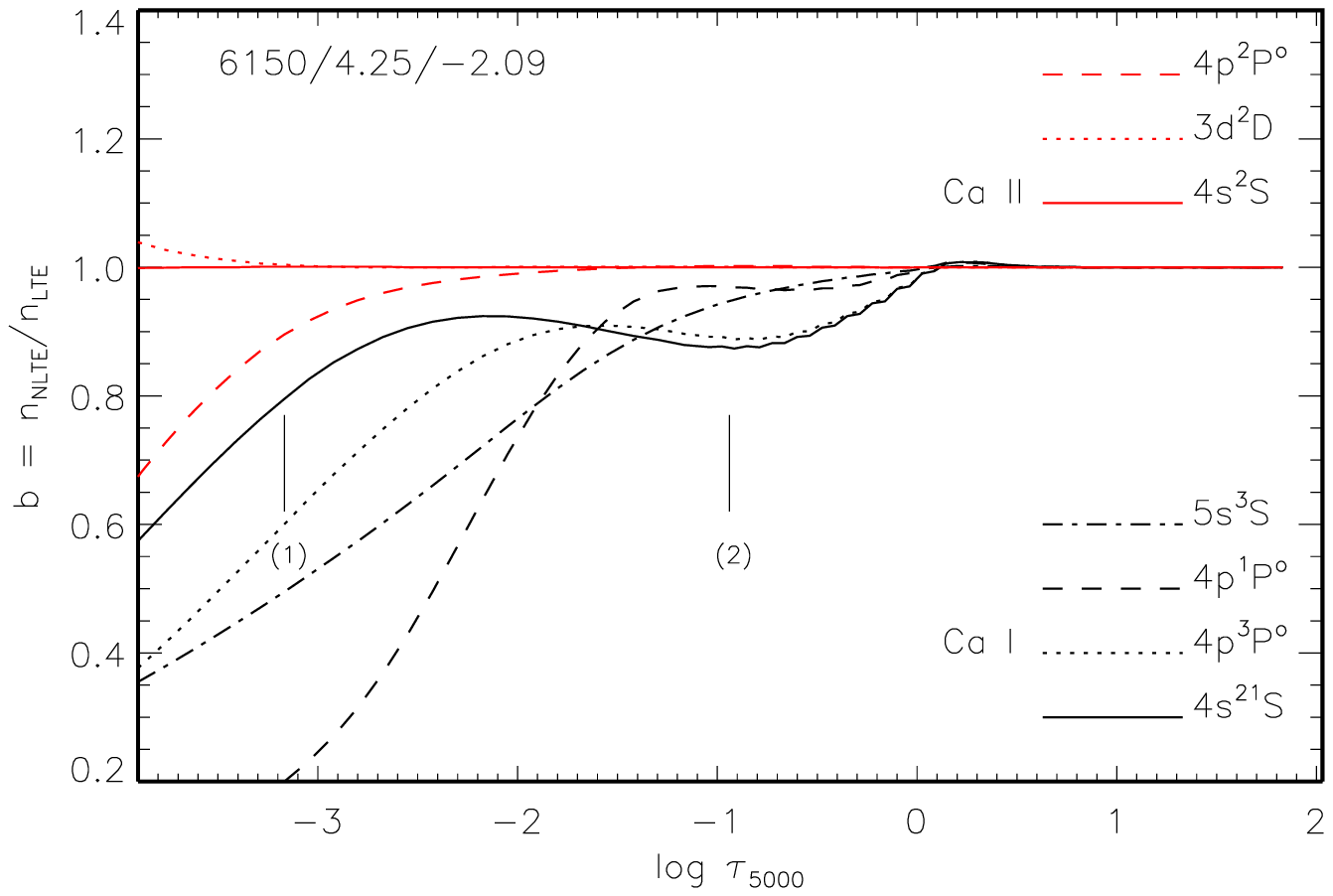}}
\resizebox{60mm}{!}{\includegraphics{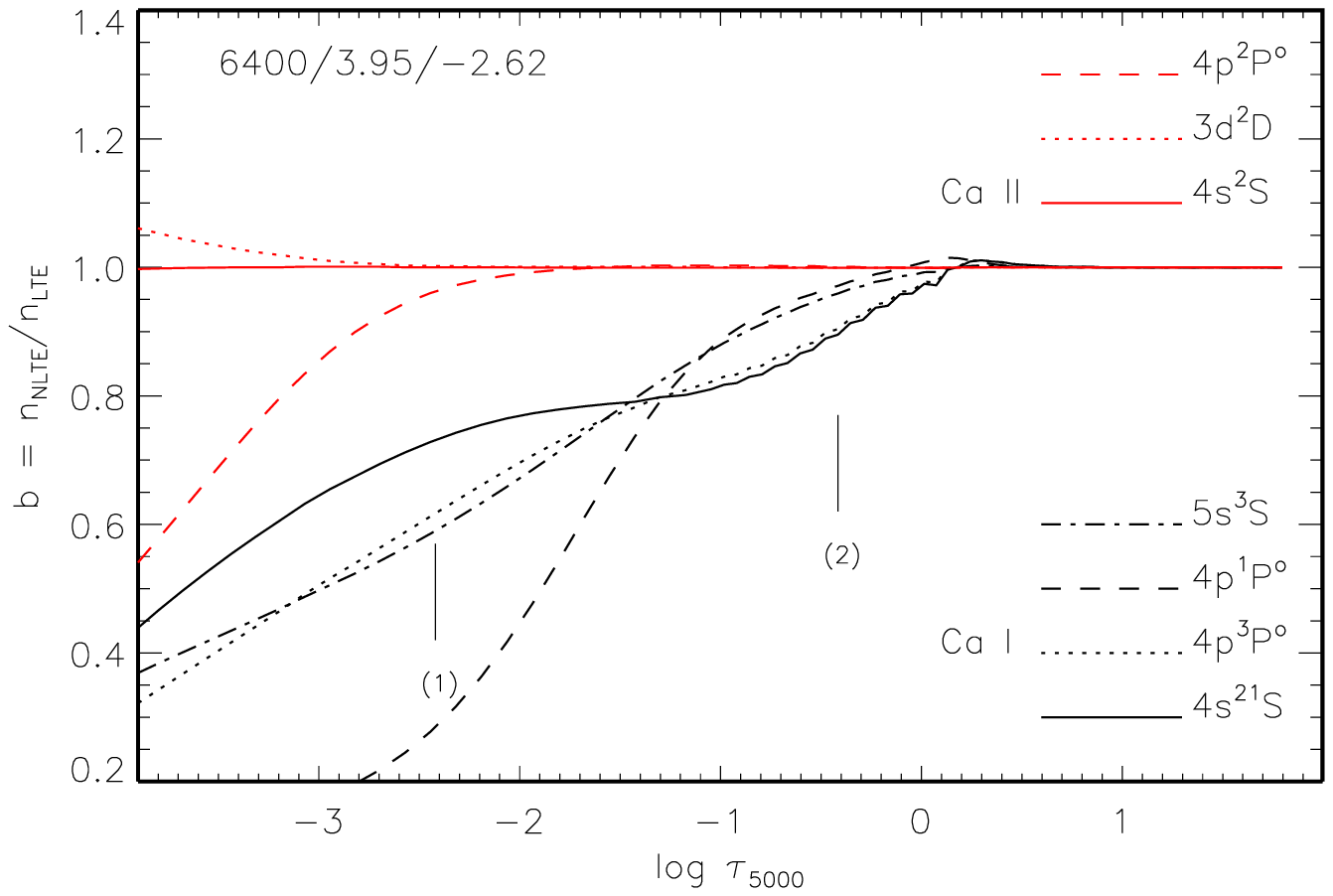}}
\resizebox{60mm}{!}{\includegraphics{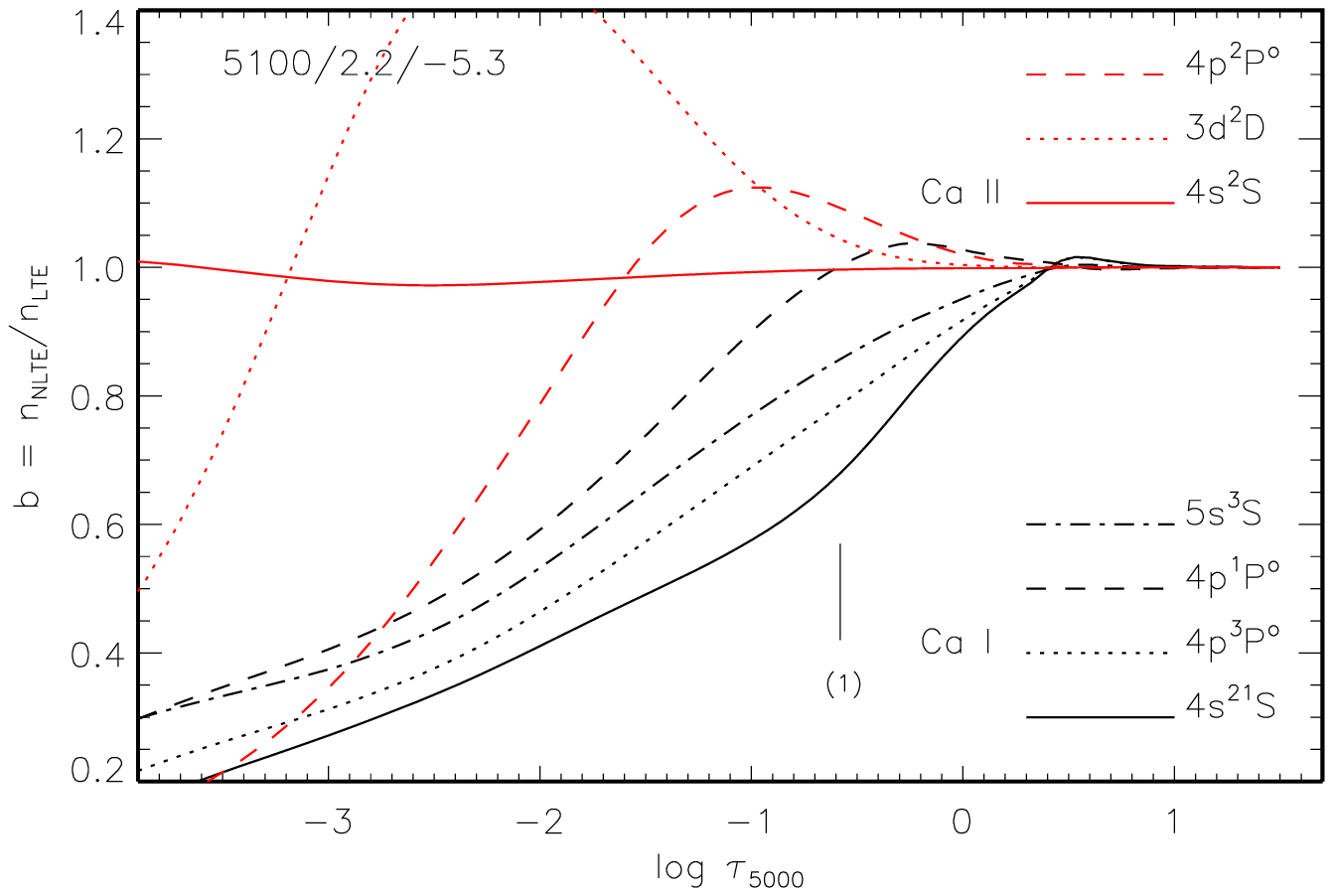}}

\vspace{-1mm}
\resizebox{60mm}{!}{\includegraphics{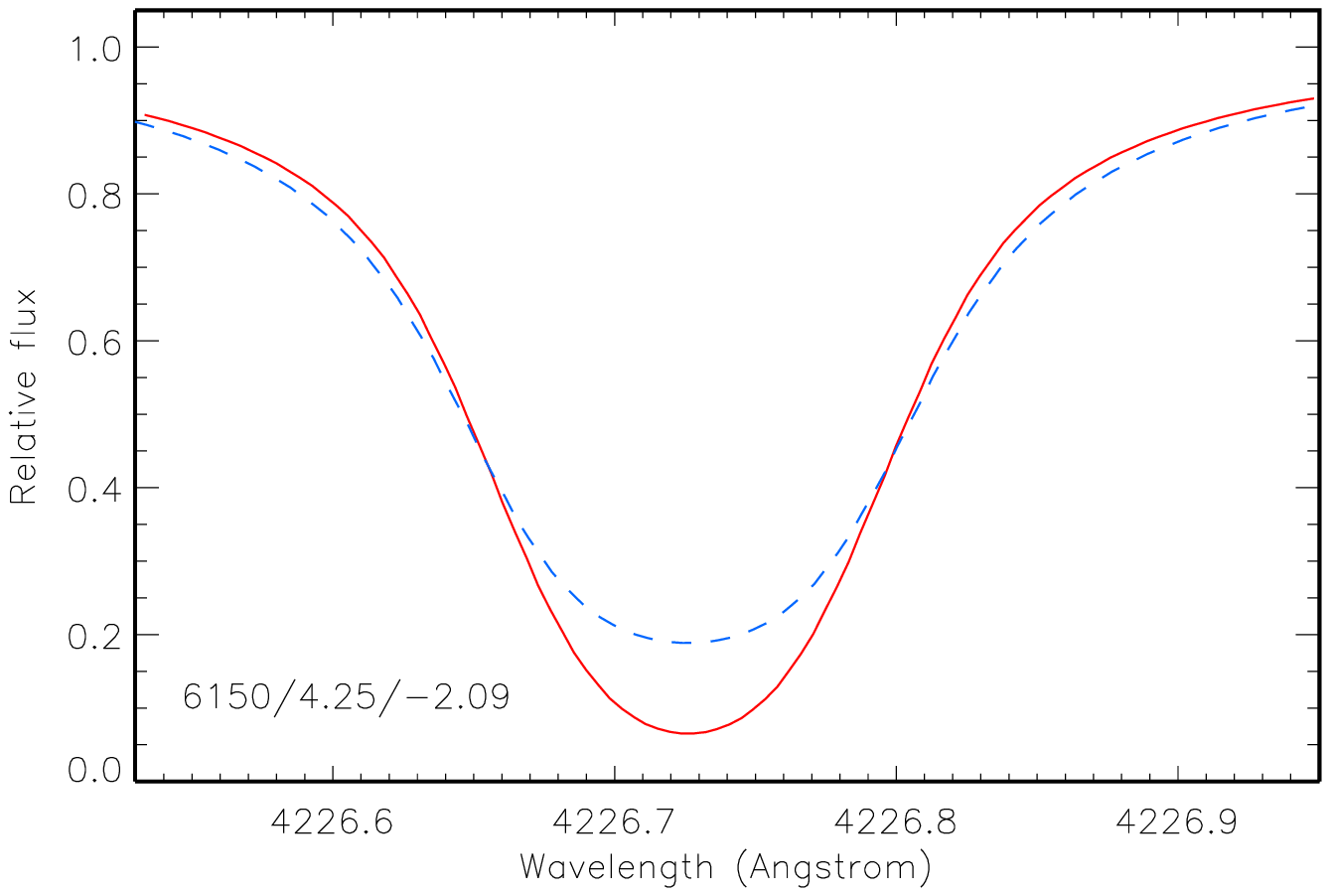}}
\resizebox{60mm}{!}{\includegraphics{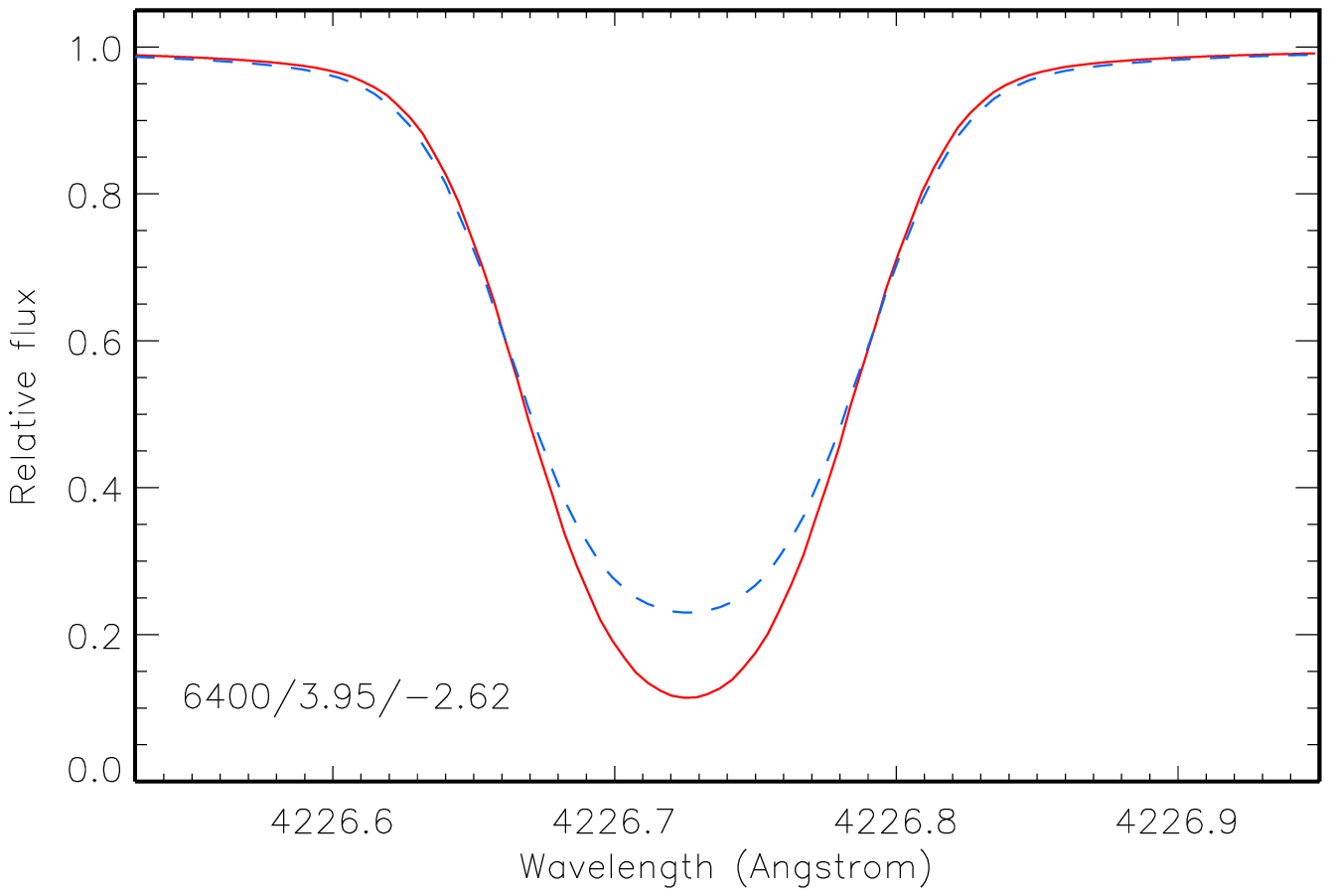}}
\resizebox{60mm}{!}{\includegraphics{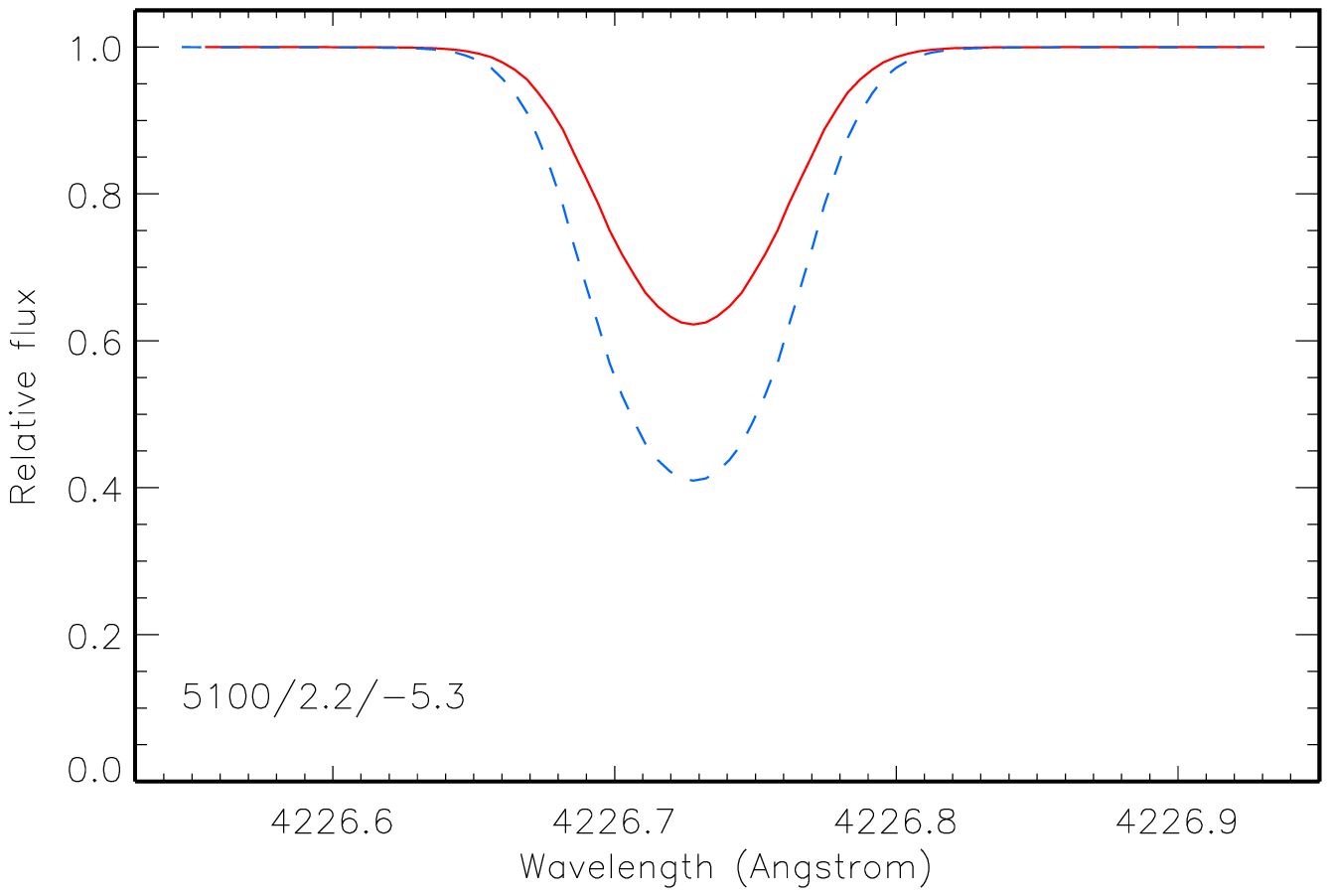}}
\caption[]{Top row: departure coefficients for selected levels of \ion{Ca}{i} and \ion{Ca}{ii} as a function of $\log \tau_{5000}$ in the models representing atmospheres of BD~$+09^\circ$0352, BD~$-13^\circ$3442, and HE~0107-5240. Tick marks indicate the locations of line centre optical depth unity for \ion{Ca}{i} 4226\,\AA\ (1) and \ion{Ca}{i} 6162\,\AA\ (2). Bottom row: NLTE (continuous curve) and LTE (dashed curve) theoretical profiles of \ion{Ca}{i} 4226\,\AA\ in the corresponding models. }  \label{Fig:ca4226}
\end{figure*}

To clarify the situation with \ion{Ca}{i} 4226\,\AA, we plot in Fig.\,\ref{Fig:ca4226} the departure coefficients of the lower (\eu{4s}{1}{S}{}{}) and upper (\eu{4p}{1}{P}{\circ}{}) levels of the corresponding transition in three model atmospheres of different metallicity. We chose 
BD$+09^\circ$0352 ([Fe/H] = $-2.09$), BD$-13^\circ$3442 ([Fe/H] = $-2.62$), and HE~0107-5240 ([Fe/H] = $-5.3$).
For comparison, departure coefficients of the lower and upper levels of \ion{Ca}{i} 6162\,\AA\ (\eu{4p}{3}{P}{\circ}{} -- \eu{5s}{3}{S}{}{}) are also shown. In the [Fe/H] = $-5.3$ model, \ion{Ca}{i} 4226\,\AA\ forms in deep atmospheric layers where the overall over-ionisation of \ion{Ca}{i} leads to a weakened line (bottom row of Fig.\,\ref{Fig:ca4226}) and positive NLTE abundance correction (Fig.\,\ref{Fig:ca4226_dnlte}). In contrast, in the [Fe/H] = $-2.09$ model, the core of the resonance line forms in the uppermost atmospheric layers, where b(\eu{4p}{1}{P}{\circ}{}) $<$ b(\eu{4s}{1}{S}{}{}) due to photon loss in the line itself and the line source function is smaller than the Planck function, resulting in enhanced absorption in the line core. However, at this metallicity, the line has broad wings that form in deep atmospheric layers and are weakened by over-ionisation of \ion{Ca}{i}. This compensates, in part, for the NLTE effects in the line core and results in an only slightly negative NLTE abundance correction of $\Delta_{\rm NLTE} = -0.04$~dex. In the [Fe/H] = $-2.62$ model, \ion{Ca}{i} 4226\,\AA\ nearly looses its line wings, and enhanced absorption in the line core determines the NLTE effects on the total strength of the line. 

We thus understand why in a certain range of the Ca abundance, NLTE modelling leads to a too strong \ion{Ca}{i} 4226\,\AA\ line and to a too low abundance derived from this line, but we have no tool to influence the SE of \ion{Ca}{i} in the uppermost atmospheric layers where the UV over-ionisation and photon escapes in the resonance line decide the situation. 

Based on this analysis, we suggest that NLTE modelling provides reliable results for \ion{Ca}{i} 4226\,\AA\ when the calcium abundance is low, [Ca/H] $\precsim -5$, and the \ion{Ca}{i}/\ion{Ca}{ii} ionisation equilibrium method can successfully be applied to determine surface gravities of the hyper metal-poor stars.
For example, if the abundance difference \ion{Ca}{i} -- \ion{Ca}{ii} is accurate within 0.1~dex, the uncertainty in log~g of HMP turn-off star does not exceed 0.2~dex.

\section{NLTE abundance corrections for lines of \ion{Ca}{i} depending on stellar parameters} \label{sect:corrections}

\begin{figure*}
\resizebox{67mm}{!}{\includegraphics{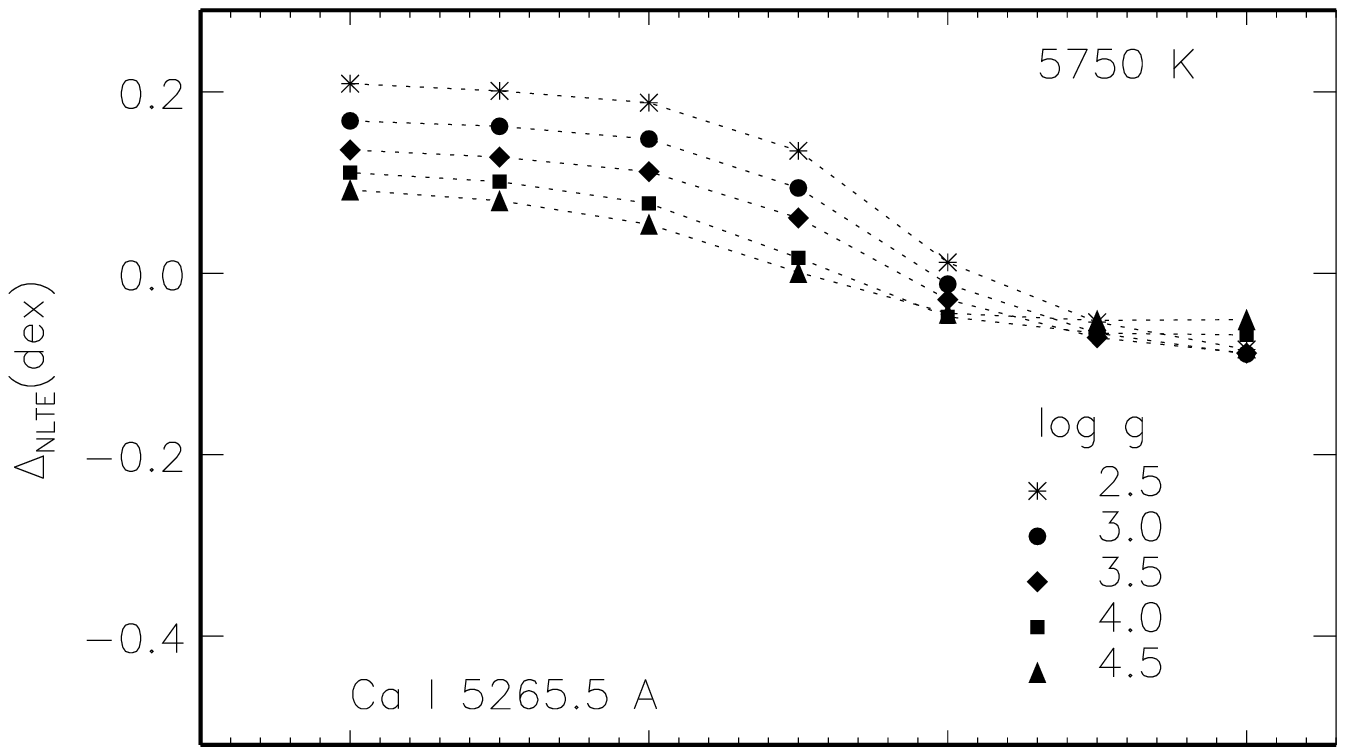}}
\hspace{-16.5mm}
\resizebox{67mm}{!}{\includegraphics{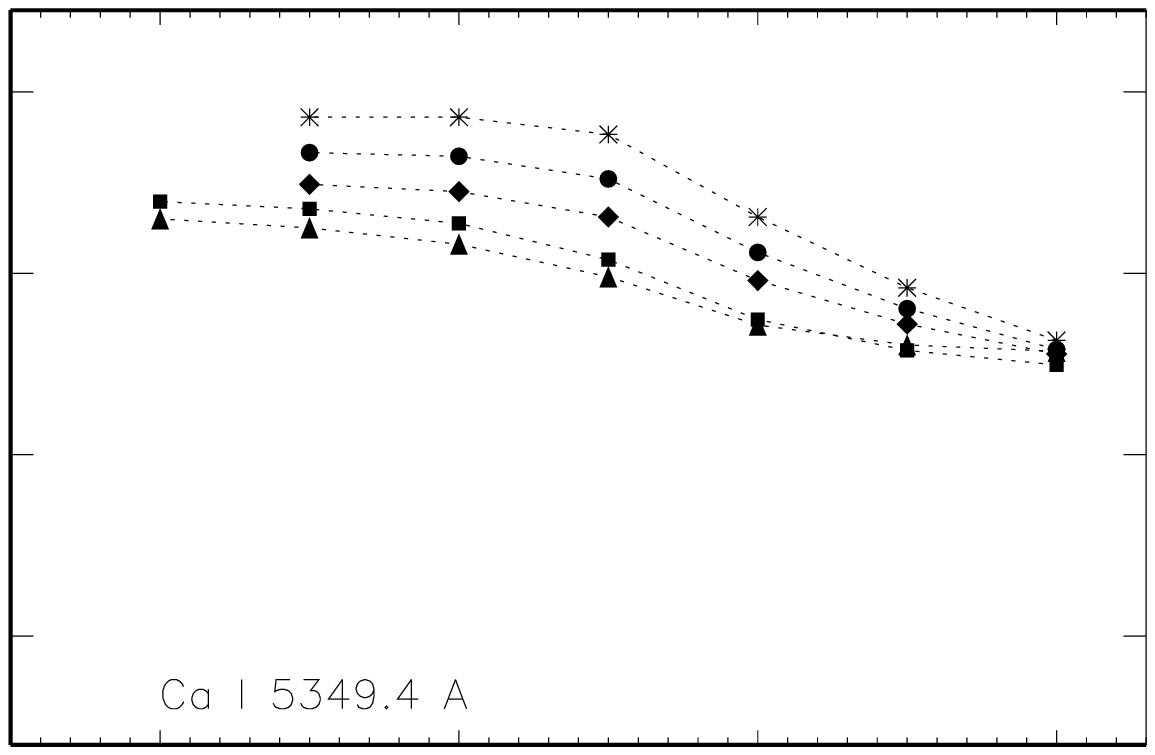}}
\hspace{-16.5mm}
\resizebox{67mm}{!}{\includegraphics{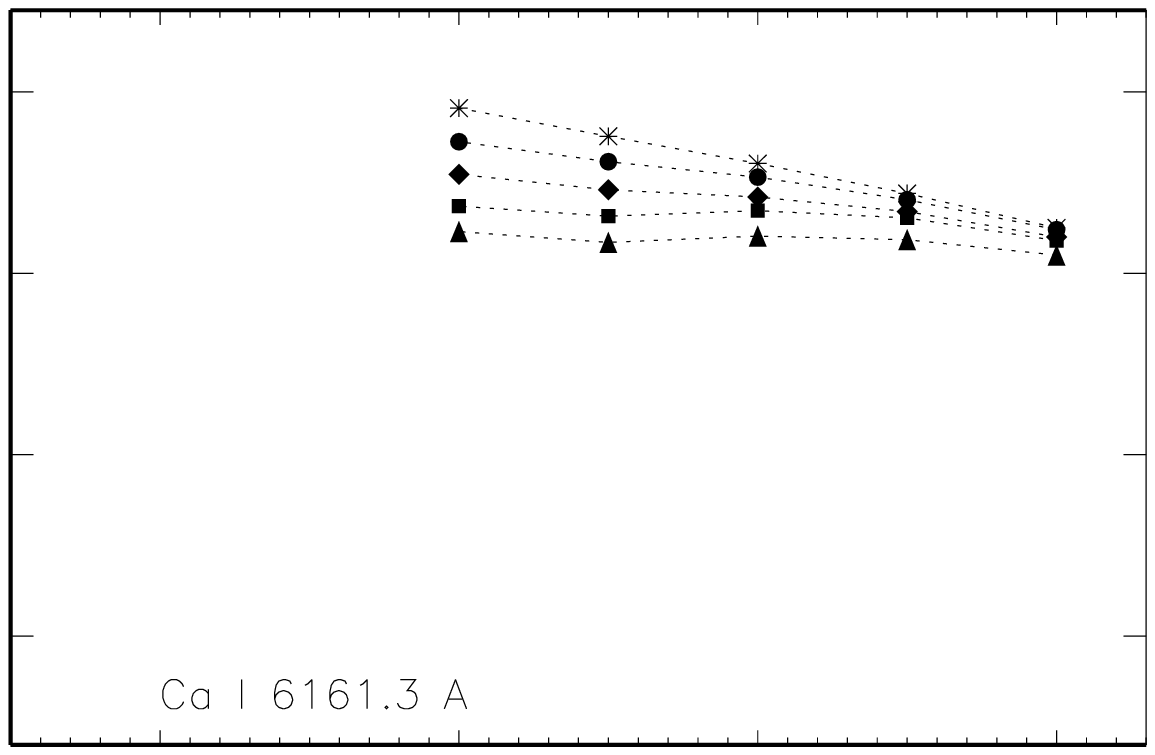}}

\vspace{-11mm}
\resizebox{67mm}{!}{\includegraphics{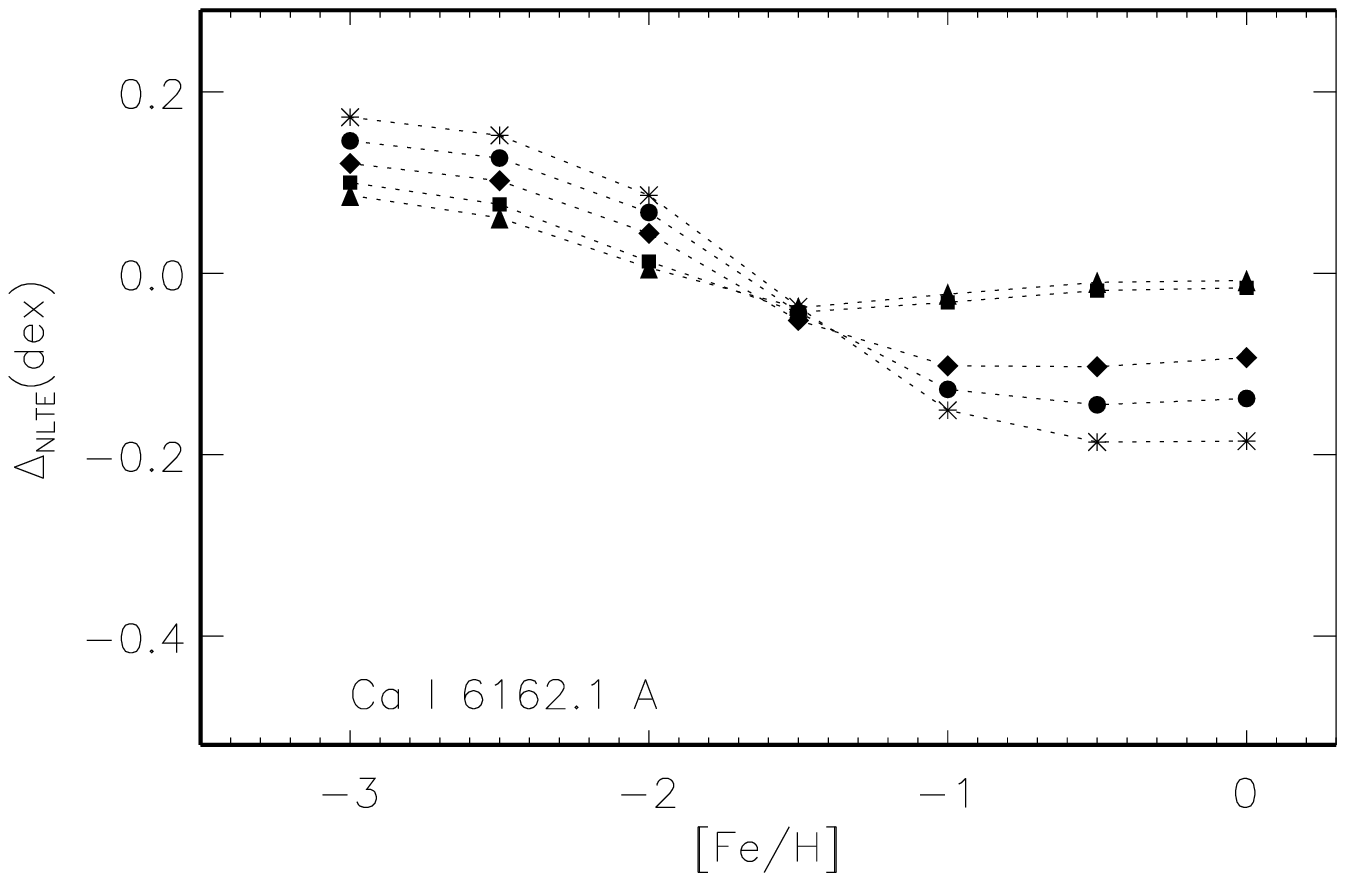}}
\hspace{-16.5mm}
\resizebox{67mm}{!}{\includegraphics{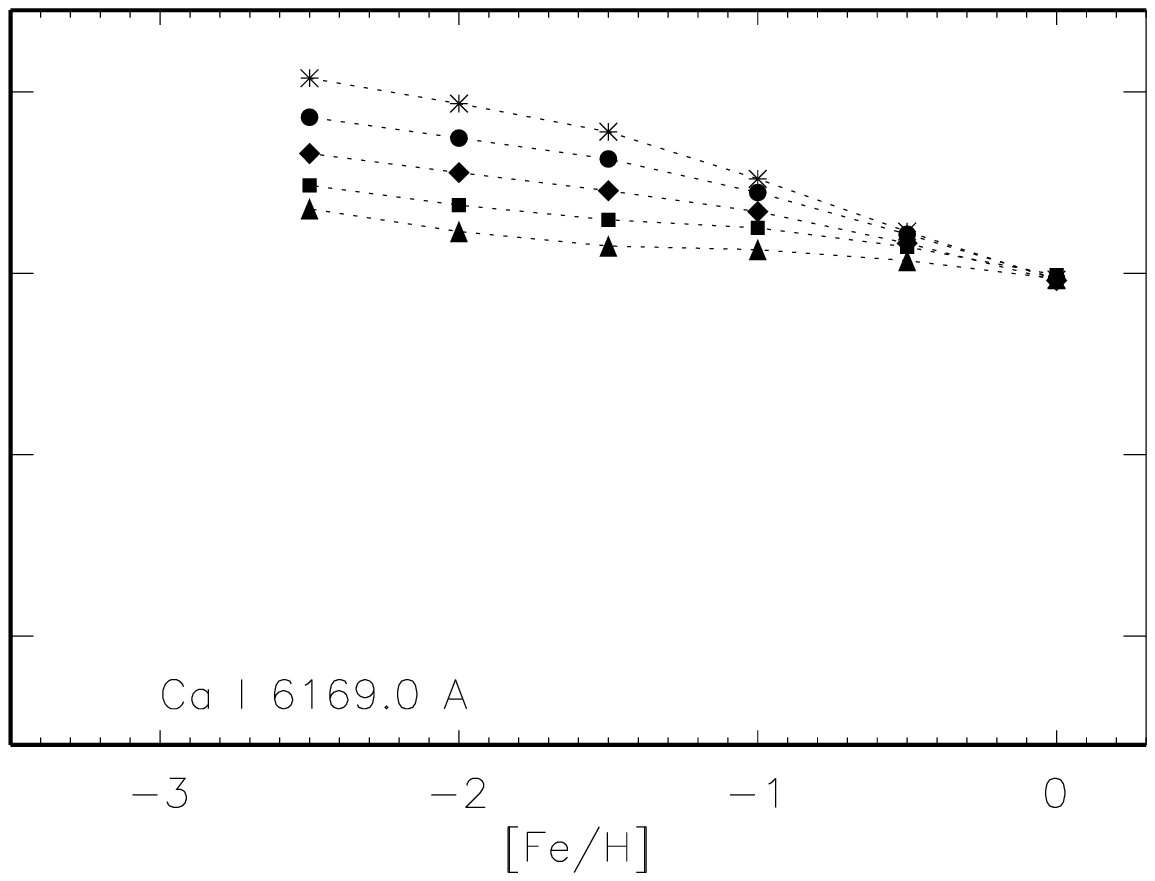}}
\hspace{-16.5mm}
\resizebox{67mm}{!}{\includegraphics{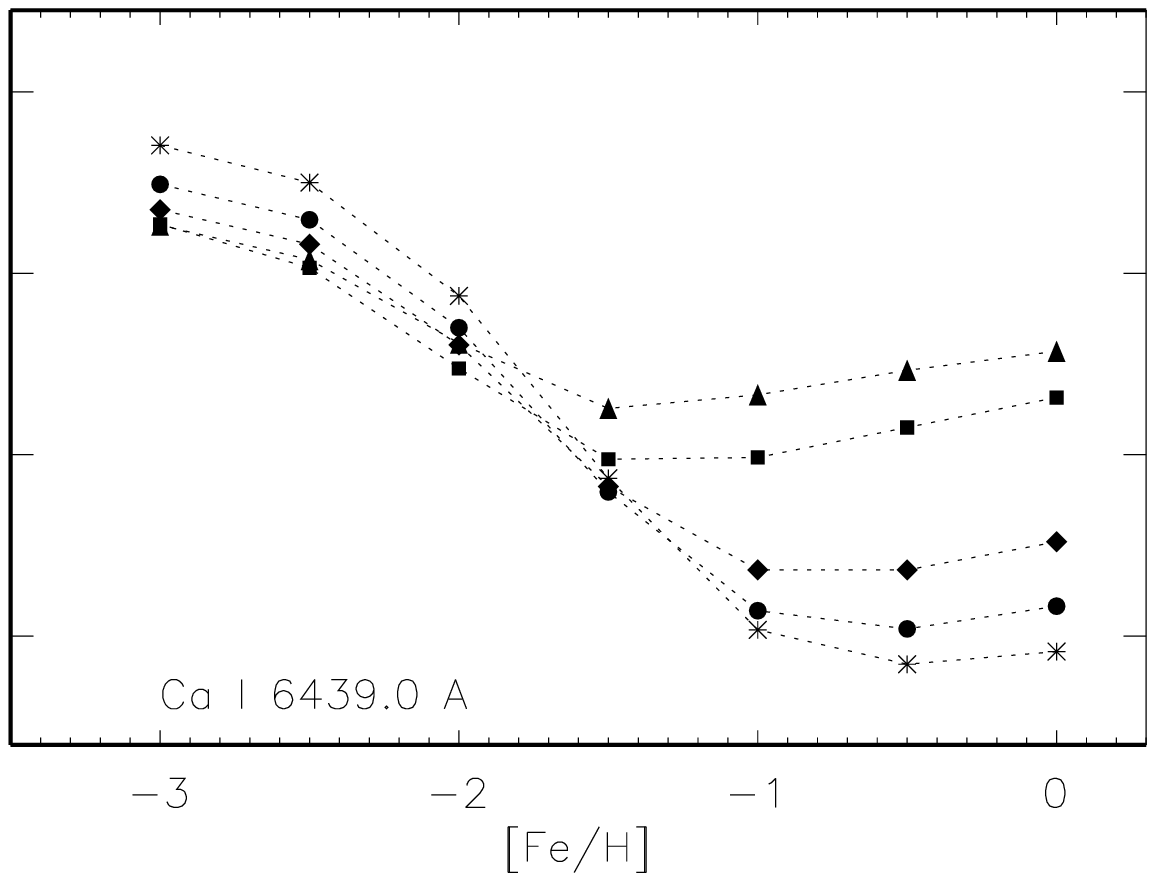}}
\caption[]{NLTE (MGB17) abundance corrections for selected lines of \ion{Ca}{i} depending on metallicity and surface gravity in the models with common $\Teff$ = 5750~K. Different symbols correspond to different surface gravities: log~g = 2.5 (asterisks), 3.0 (circles), 3.5 (rhombi), 4.0 (squares), and 4.5 (triangles).}  \label{Fig:dnlte}
\end{figure*}

With the MGB17 collisional data, we computed the NLTE abundance corrections for 28 lines of \ion{Ca}{i} in the grid of MARCS model atmospheres, where an effective temperature varies between 5\,000~K and 6\,500~K, with a 250~K step, and surface gravity between log~g = 2.5 and 4.5, with a 0.5 step. The metallicity ranges between [Fe/H] = 0 and $-4$. In part, the obtained results are displayed in Fig.\,\ref{Fig:dnlte}. The entire data set is available in machine-readable form at the site {\tt http://www.inasan.ru/$\sim$lima/}.

Comments on Fig.\,\ref{Fig:dnlte}. 

-- The NLTE corrections are not provided for model atmospheres where a given line is weak, with an equivalent width smaller than 3\,m\AA. Even the strongest subordinate lines of \ion{Ca}{i} are weak in the [Fe/H] = $-4$ models.

-- A common feature of most \ion{Ca}{i} lines is the change in sign of the NLTE abundance correction when moving from the solar metallicity and mildly metal-poor model atmospheres, where $\Delta_{\rm NLTE} < 0$, to the very metal-poor ones, where $\Delta_{\rm NLTE} > 0$. When a line is strong, its wings are weakened, but the core is stronger than in the LTE case. The value and sign of $\Delta_{\rm NLTE}$ are defined by a relative contribution of the core and the wings to the overall line strength. When the line becomes weak as a result of the decreasing Ca abundance, NLTE leads to depleted total absorption in the line and to a positive abundance correction. 
The metallicity at which a change in sign occurs is different for different lines, and for a given line, it  depends on atmospheric $\Teff$ and log~g. In the $\Teff$ = 5750~K models, $\Delta_{\rm NLTE}$ is only positive throughout the investigated metallicity range for \ion{Ca}{i} 6161 and 6169.0\,\AA.

-- In general, the NLTE effects grow towards lower surface gravity. For \ion{Ca}{i} 6161 and 6169.0\,\AA, $\Delta_{\rm NLTE}$ grows towards lower metallicity in the entire metallicity range and, for the remaining lines, at [Fe/H] $< -1.5$. 

It is worth noting that the NLTE abundance corrections for lines of \ion{Ca}{i} in the cool giant model atmospheres (4000~K $\le \Teff \le$ 5000~K, 0.5 $\le$ log~g $\le$ 2.5, $-4 \le$ [Fe/H] $\le 0$) were computed by \citet{Mashonkina_dnlte2016} using rate coefficients for \ion{Ca}{i} + \ion{H}{i} collisions from quantum-mechanical calculations of \citet{Belyaev2016_Ca}. The data are accessible online via {\tt http://spectrum.inasan.ru/nLTE/}, and can be downloaded in machine-readable form at the site {\tt http://www.inasan.ru/$\sim$lima/}. We have inspected differences in $\Delta_{\rm NLTE}$ between applying data of \citet{Belyaev2016_Ca} and data from \citet{ca1_hydrogen}. They are minor in most cases. For example, in the 4500/1.5/$-3$ model, the NLTE abundance corrections based on the MGB17 data are smaller, by 0.02-0.03~dex, for the $W \precsim$ 30\,m\AA\ lines, but larger, by 0.02-0.03~dex, for the stronger lines compared with the corresponding values computed with the rate coefficients from \citet{Belyaev2016_Ca}.

\section{Conclusions and recommendations} \label{conclusion}

We performed the NLTE calculations for \ion{Ca}{i-ii} with the updated model atom that includes inelastic \ion{Ca}{i}+\ion{H}{i} collisions using accurate rate coefficients from quantum-mechanical calculations of \citet{ca1_hydrogen}. For \ion{Ca}{ii}, we still applied the classical Drawinian rates scaled by \kH\ = 0.1. For the model atmosphere representative of the VMP turnoff stars, 5780/3.7/$-2.46$, we showed that including \ion{H}{i} collisions substantially reduces the over-ionisation of \ion{Ca}{i} in the line formation layers compared with the case of pure electronic collisions, such that the NLTE abundance corrections become smaller, by 0.08 to 0.20~dex, for different \ion{Ca}{i} lines. The calculations of \citet{ca1_hydrogen} and \citet[][updated]{2016PhRvA..93d2705B} provide very similar rate coefficients for \ion{Ca}{i}+\ion{H}{i} collisions.

The NLTE abundances based on the MGB17 collisional data were derived from the \ion{Ca}{i} subordinate lines in the Sun, Procyon, and five MP stars in the $-2.62 \le$ [Fe/H] $\le -1.26$ metallicity range.
In order to inspect the \ion{Ca}{i}/\ion{Ca}{ii} ionisation equilibrium with given atmospheric parameters, we determined abundances from the high-excitation lines of \ion{Ca}{ii}, and for the MP stars, also from \ion{Ca}{ii} 8498\,\AA.  
The analysis of 22 solar \ion{Ca}{i} lines with the MARCS solar model atmosphere yields the NLTE abundance $\eps{CaI}$ = 6.33 ($\sigma$ = 0.06~dex), in line with the meteoritic value of \citet{Lodders2009} and the photospheric abundance based on the 3D+NLTE calculations of \citet{2015A&A...573A..25S}. The NLTE abundance calculated from solar high-excitation lines of \ion{Ca}{ii}, $\eps{CaII}$ = 6.40 ($\sigma$ = 0.05~dex), is higher than that from the \ion{Ca}{ii} 8498\,\AA\ line wing fit, by 0.13~dex. For the stars, we determined the differential NLTE and
LTE abundances relative to the Sun. The obtained results can be summarised as follows.
\begin{itemize}
\item For both \ion{Ca}{i} and \ion{Ca}{ii} in every star including the Sun, NLTE leads to smaller line-to-line scatter, such that $\sigma$ reduces compared with the LTE one, by up to a factor of 2.5.
\item For Procyon, NLTE removes a steep trend with line strength among strong \ion{Ca}{i} lines seen in LTE.
\item For Procyon, we achieve NLTE abundances consistent within 0.06~dex from the two ionisation stages, \ion{Ca}{i} and \ion{Ca}{ii}.
\item In each of the five MP stars, \ion{Ca}{ii} 8498\,\AA\ yields an NLTE abundance that agrees well with that from the \ion{Ca}{i} subordinate lines. For comparison, the difference in LTE abundances grows towards lower metallicity and reaches 0.46~dex in BD~$-13^\circ$3442 ([Fe/] = $-2.62$).
\item For the four VMP stars, NLTE largely removes obvious abundance discrepancies between the high-excitation lines of \ion{Ca}{ii} and \ion{Ca}{ii} 8498\,\AA\ obtained under LTE assumption.
\end{itemize}

We thus strongly recommend to apply the NLTE approach to Ca abundance determinations. This is important for any individual star because the magnitude and sign of the NLTE abundance correction can be different for different lines of \ion{Ca}{i} and \ion{Ca}{ii} and for stellar samples covering wide metallicity ranges. The SE equilibrium calculations have to include inelastic \ion{Ca}{i} + \ion{H}{i} collisions.

With the updated model atom, we investigate the formation of the \ion{Ca}{i} 4226\,\AA\ resonance line in the [Fe/H] $< -2$ stars, where it has a purely photospheric origin. In the four stars with $-2.3 <$ [Ca/H] $< -1.8$ ($-2.62 \le$ [Fe/H] $\le -2.09$), \ion{Ca}{i} 4226\,\AA\ is strong enough, with $W_{obs} \ge$ 113\,m\AA, and NLTE predicts a weakening of the line wings and strengthening of the line core compared with the LTE case, resulting in negative NLTE abundance correction. A relative contribution of the core to the overall line strength grows with decreasing Ca abundance, and $\Delta_{\rm NLTE}$ increases in absolute value. In these stars, the \ion{Ca}{i} resonance line yields a lower abundance than the abundance from the subordinate lines, by 0.08 to 0.32~dex.
We can explain why the abundance derived from \ion{Ca}{i} 4226\,\AA\ is underestimated, but cannot influence the photon loss in the \ion{Ca}{i} resonance line in the uppermost atmospheric layers that is the driver of strengthening the line core, which results in a negative NLTE abundance correction. 

When we decrease the Ca abundance to [Ca/H] $\simeq -4.4$, like that of the UMP star HE~0557-4840, the line core formation depth shifts to deep atmospheric layers, where over-ionisation of \ion{Ca}{i} determines the  depleted total absorption in the line and a positive abundance correction. 
Nevertheless, the NLTE abundance obtained from \ion{Ca}{i} 4226\,\AA\ in HE~0557-4840 is lower than that from five \ion{Ca}{ii} lines, by 0.26~dex.

The NLTE effects for \ion{Ca}{i} 4226\,\AA\ grow towards lower Ca abundance, and we computed large positive NLTE corrections of $\Delta_{\rm NLTE}$ = 0.34 and 0.25~dex for the two HMP stars HE~0107-5240 and HE~1327-2326, respectively. For each of them, the NLTE abundances from the \ion{Ca}{i} resonance line and the \ion{Ca}{ii} lines are found to be consistent.
Thus, the \ion{Ca}{i}/\ion{Ca}{ii} ionisation equilibrium method can successfully be applied to determine surface gravities of hyper metal-poor stars.

\begin{acknowledgements}
This study is based on observations made with ESO Telescopes at the La Silla Paranal Observatory under programme IDs 075.D-0048(A), 077.D-0035(A), 0.D-0009(A), 076.D-0165(A), and 380.D-0040(A) 
 and with the Canada-France-Hawaii Telescope under programme IDs 1424029 and 1515097-98. This research used the services of the ESO Science Archive Facility and the facilities of the Canadian Astronomy Data Centre operated by the National Research Council of Canada with the support of the Canadian Space Agency. 
The authors acknowledge financial support from the Russian Scientific Foundation (grant 17-13-01144). We made use the MARCS and VALD databases. We thank Oleg Kochukhov for providing the
{\sc idl binmag}3 code.
\end{acknowledgements}

\bibliography{mashonkina,atomic_data,nlte,mp_stars,references}

\begin{thebibliography}{72}
\expandafter\ifx\csname natexlab\endcsname\relax\def\natexlab#1{#1}\fi

\bibitem[{{Andretta} {et~al.}(2005){Andretta}, {Bus{\`a}}, {Gomez}, \&
  {Terranegra}}]{2005A&A...430..669A}
{Andretta}, V., {Bus{\`a}}, I., {Gomez}, M.~T., \& {Terranegra}, L. 2005, \aap,
  430, 669

\bibitem[{{Anstee} \& {O'Mara}(1995)}]{1995MNRAS.276..859A}
{Anstee}, S.~D. \& {O'Mara}, B.~J. 1995, \mnras, 276, 859

\bibitem[{{Aoki} {et~al.}(2006){Aoki}, {Frebel}, {Christlieb}, {Norris},
  {Beers}, {Minezaki}, {Barklem}, {Honda}, {Takada-Hidai}, {Asplund}, {Ryan},
  {Tsangarides}, {Eriksson}, {Steinhauer}, {Deliyannis}, {Nomoto}, {Fujimoto},
  {Ando}, {Yoshii}, \& {Kajino}}]{Aoki_he1327}
{Aoki}, W., {Frebel}, A., {Christlieb}, N., {et~al.} 2006, \apj, 639, 897

\bibitem[{{Bagnulo} {et~al.}(2003){Bagnulo}, {Jehin}, {Ledoux}, {Cabanac},
  {Melo}, {Gilmozzi}, \& {ESO Paranal Science Operations
  Team}}]{2003Msngr.114...10B}
{Bagnulo}, S., {Jehin}, E., {Ledoux}, C., {et~al.} 2003, The Messenger, 114, 10

\bibitem[{{Barklem}(2016)}]{2016PhRvA..93d2705B}
{Barklem}, P.~S. 2016, \pra, 93, 042705

\bibitem[{{Barklem} \& {O'Mara}(1997)}]{1997MNRAS.290..102B}
{Barklem}, P.~S. \& {O'Mara}, B.~J. 1997, \mnras, 290, 102

\bibitem[{{Barklem} \& {O'Mara}(1998)}]{1998MNRAS.300..863B}
{Barklem}, P.~S. \& {O'Mara}, B.~J. 1998, \mnras, 300, 863

\bibitem[{{Barklem} {et~al.}(1998){Barklem}, {O'Mara}, \&
  {Ross}}]{1998MNRAS.296.1057B}
{Barklem}, P.~S., {O'Mara}, B.~J., \& {Ross}, J.~E. 1998, \mnras, 296, 1057

\bibitem[{{Battaglia} {et~al.}(2008){Battaglia}, {Helmi}, {Tolstoy}, {Irwin},
  {Hill}, \& {Jablonka}}]{2008ApJ...681L..13B}
{Battaglia}, G., {Helmi}, A., {Tolstoy}, E., {et~al.} 2008, \apjl, 681, L13

\bibitem[{{Belyaev} {et~al.}(2016){Belyaev}, {Yakovleva}, {Guitou},
  {Mitrushchenkov}, {Spielfiedel}, \& {Feautrier}}]{Belyaev2016_Ca}
{Belyaev}, A.~K., {Yakovleva}, S.~A., {Guitou}, M., {et~al.} 2016, \aap, 587,
  A114

\bibitem[{{Bond} {et~al.}(2015){Bond}, {Gilliland}, {Schaefer}, {Demarque},
  {Girard}, {Holberg}, {Gudehus}, {Mason}, {Kozhurina-Platais}, {Burleigh},
  {Barstow}, \& {Nelan}}]{2015ApJ...813..106B}
{Bond}, H.~E., {Gilliland}, R.~L., {Schaefer}, G.~H., {et~al.} 2015, \apj, 813,
  106

\bibitem[{{Bonifacio} {et~al.}(2015){Bonifacio}, {Caffau}, {Spite}, {Limongi},
  {Chieffi}, {Klessen}, {Fran{\c c}ois}, {Molaro}, {Ludwig}, {Zaggia}, {Spite},
  {Plez}, {Cayrel}, {Christlieb}, {Clark}, {Glover}, {Hammer}, {Koch},
  {Monaco}, {Sbordone}, \& {Steffen}}]{2015A&A...579A..28B}
{Bonifacio}, P., {Caffau}, E., {Spite}, M., {et~al.} 2015, \aap, 579, A28

\bibitem[{{Boyajian} {et~al.}(2013){Boyajian}, {von Braun}, {van Belle},
  {Farrington}, {Schaefer}, {Jones}, {White}, {McAlister}, {ten Brummelaar},
  {Ridgway}, {Gies}, {Sturmann}, {Sturmann}, {Turner}, {Goldfinger}, \&
  {Vargas}}]{Boyajian2013}
{Boyajian}, T.~S., {von Braun}, K., {van Belle}, G., {et~al.} 2013, \apj, 771,
  40

\bibitem[{{Butler} \& {Giddings}(1985)}]{detail}
{Butler}, K. \& {Giddings}, J. 1985, Newsletter on the analysis of astronomical
  spectra, No. 9, University of London

\bibitem[{{Caffau} {et~al.}(2011){Caffau}, {Bonifacio}, {Fran{\c c}ois},
  {Sbordone}, {Monaco}, {Spite}, {Spite}, {Ludwig}, {Cayrel}, {Zaggia},
  {Hammer}, {Randich}, {Molaro}, \& {Hill}}]{2011Natur.477...67C}
{Caffau}, E., {Bonifacio}, P., {Fran{\c c}ois}, P., {et~al.} 2011, \nat, 477,
  67

\bibitem[{{Caffau} {et~al.}(2012){Caffau}, {Bonifacio}, {Fran{\c c}ois},
  {Spite}, {Spite}, {Zaggia}, {Ludwig}, {Steffen}, {Mashonkina}, {Monaco},
  {Sbordone}, {Molaro}, {Cayrel}, {Plez}, {Hill}, {Hammer}, \&
  {Randich}}]{2012A&A...542A..51C}
{Caffau}, E., {Bonifacio}, P., {Fran{\c c}ois}, P., {et~al.} 2012, \aap, 542,
  A51

\bibitem[{{Cayrel} {et~al.}(2004){Cayrel}, {Depagne}, {Spite}, {Hill}, {Spite},
  {Fran{\c c}ois}, {Plez}, {Beers}, {Primas}, {Andersen}, {Barbuy},
  {Bonifacio}, {Molaro}, \& {Nordstr{\"o}m}}]{Cayrel2004}
{Cayrel}, R., {Depagne}, E., {Spite}, M., {et~al.} 2004, \aap, 416, 1117

\bibitem[{{Chiavassa} {et~al.}(2012){Chiavassa}, {Bigot}, {Kervella}, {Matter},
  {Lopez}, {Collet}, {Magic}, \& {Asplund}}]{2012A&A...540A...5C}
{Chiavassa}, A., {Bigot}, L., {Kervella}, P., {et~al.} 2012, \aap, 540, A5

\bibitem[{{Christlieb} {et~al.}(2002){Christlieb}, {Bessell}, {Beers},
  {Gustafsson}, {Korn}, {Barklem}, {Karlsson}, {Mizuno-Wiedner}, \&
  {Rossi}}]{2002Natur.419..904C}
{Christlieb}, N., {Bessell}, M.~S., {Beers}, T.~C., {et~al.} 2002, \nat, 419,
  904

\bibitem[{Christlieb {et~al.}(2004)Christlieb, Gustafsson, Korn, Barklem,
  Beers, Bessell, Karlsson, \& Mizuno-Wiedner}]{HE0107_ApJ}
Christlieb, N., Gustafsson, B., Korn, A., {et~al.} 2004, ApJ, 603, 708

\bibitem[{{Cohen} {et~al.}(2013){Cohen}, {Christlieb}, {Thompson}, {McWilliam},
  {Shectman}, {Reimers}, {Wisotzki}, \& {Kirby}}]{Cohen2013}
{Cohen}, J.~G., {Christlieb}, N., {Thompson}, I., {et~al.} 2013, \apj, 778, 56

\bibitem[{{Drake}(1991)}]{1991MNRAS.251..369D}
{Drake}, J.~J. 1991, \mnras, 251, 369

\bibitem[{{Drawin}(1968)}]{Drawin1968}
{Drawin}, H.-W. 1968, Zeitschrift fur Physik, 211, 404

\bibitem[{{Drawin}(1969)}]{Drawin1969}
{Drawin}, H.~W. 1969, Zeitschrift fur Physik, 225, 483

\bibitem[{{Drozdowski} {et~al.}(1997){Drozdowski}, {Ignaciuk}, {Kwela}, \&
  {Heldt}}]{1997ZPhyD..41..125D}
{Drozdowski}, R., {Ignaciuk}, M., {Kwela}, J., \& {Heldt}, J. 1997, Zeitschrift
  fur Physik D Atoms Molecules Clusters, 41, 125

\bibitem[{Frebel {et~al.}(2005)Frebel, Aoki, Christlieb, Ando, Asplund,
  Barklem, Beers, Eriksson, Fechner, Fujimoto, Honda, Kajino, Minezaki, Nomoto,
  Norris, Ryan, Takada-Hidai, Tsangarides, \& Yoshii}]{Frebeletal:2005}
Frebel, A., Aoki, W., Christlieb, N., {et~al.} 2005, Nature, 434, 871

\bibitem[{{Frebel} {et~al.}(2015){Frebel}, {Chiti}, {Ji}, {Jacobson}, \&
  {Placco}}]{2015ApJ...810L..27F}
{Frebel}, A., {Chiti}, A., {Ji}, A.~P., {Jacobson}, H.~R., \& {Placco}, V.~M.
  2015, \apjl, 810, L27

\bibitem[{{Fuhrmann}(1998)}]{Fuhrmann1998}
{Fuhrmann}, K. 1998, \aap, 338, 161

\bibitem[{{Fuhrmann} {et~al.}(1997){Fuhrmann}, {Pfeiffer}, {Frank}, {Reetz}, \&
  {Gehren}}]{1997A&A...323..909F}
{Fuhrmann}, K., {Pfeiffer}, M., {Frank}, C., {Reetz}, J., \& {Gehren}, T. 1997,
  \aap, 323, 909

\bibitem[{Gustafsson {et~al.}(2008)Gustafsson, Edvardsson, Eriksson, Jorgensen,
  Nordlund, \& Plez}]{Gustafssonetal:2008}
Gustafsson, B., Edvardsson, B., Eriksson, K., {et~al.} 2008, A\&A, 486, 951

\bibitem[{{Heiter} {et~al.}(2015){Heiter}, {Jofr{\'e}}, {Gustafsson}, {Korn},
  {Soubiran}, \& {Th{\'e}venin}}]{2015A&A...582A..49H}
{Heiter}, U., {Jofr{\'e}}, P., {Gustafsson}, B., {et~al.} 2015, \aap, 582, A49

\bibitem[{{Idiart} \& {Th{\'e}venin}(2000)}]{Thevenin2000}
{Idiart}, T. \& {Th{\'e}venin}, F. 2000, \apj, 541, 207

\bibitem[{{Jorgensen} {et~al.}(1992){Jorgensen}, {Carlsson}, \&
  {Johnson}}]{1992A&A...254..258J}
{Jorgensen}, U.~G., {Carlsson}, M., \& {Johnson}, H.~R. 1992, \aap, 254, 258

\bibitem[{{Keller} {et~al.}(2014){Keller}, {Bessell}, {Frebel}, {Casey},
  {Asplund}, {Jacobson}, {Lind}, {Norris}, {Yong}, {Heger}, {Magic}, {da
  Costa}, {Schmidt}, \& {Tisserand}}]{2014Natur.506..463K}
{Keller}, S.~C., {Bessell}, M.~S., {Frebel}, A., {et~al.} 2014, \nat, 506, 463

\bibitem[{{Korn} {et~al.}(2009){Korn}, {Richard}, {Mashonkina}, {Bessell},
  {Frebel}, \& {Aoki}}]{2009ApJ...698..410K}
{Korn}, A.~J., {Richard}, O., {Mashonkina}, L., {et~al.} 2009, \apj, 698, 410

\bibitem[{{Korn} {et~al.}(2003){Korn}, {Shi}, \&
  {Gehren}}]{2003A&A...407..691K}
{Korn}, A.~J., {Shi}, J., \& {Gehren}, T. 2003, \aap, 407, 691

\bibitem[{{Kurucz} {et~al.}(1984){Kurucz}, {Furenlid}, {Brault}, \&
  {Testerman}}]{Atlas}
{Kurucz}, R.~L., {Furenlid}, I., {Brault}, J., \& {Testerman}, L. 1984, {Solar
  flux atlas from 296 to 1300 nm} (New Mexico: National Solar Observatory)

\bibitem[{{Lodders} {et~al.}(2009){Lodders}, {Plame}, \& {Gail}}]{Lodders2009}
{Lodders}, K., {Plame}, H., \& {Gail}, H.-P. 2009, in Landolt-B{\"o}rnstein -
  Group VI Astronomy and Astrophysics Numerical Data and Functional
  Relationships in Science and Technology Volume 4B: Solar System. Edited by
  J.E. Tr{\"u}mper, 2009, 4.4., 44--54

\bibitem[{{Lucas} {et~al.}(2004){Lucas}, {Ramos}, {Home}, {McDonnell},
  {Nakayama}, {Stacey}, {Webster}, {Stacey}, \& {Steane}}]{2004PhRvA..69a2711L}
{Lucas}, D.~M., {Ramos}, A., {Home}, J.~P., {et~al.} 2004, \pra, 69, 012711

\bibitem[{{Mashonkina} {et~al.}(2011){Mashonkina}, {Gehren}, {Shi}, {Korn}, \&
  {Grupp}}]{mash_fe}
{Mashonkina}, L., {Gehren}, T., {Shi}, J.-R., {Korn}, A.~J., \& {Grupp}, F.
  2011, \aap, 528, A87

\bibitem[{{Mashonkina} {et~al.}(2007){Mashonkina}, {Korn}, \&
  {Przybilla}}]{mash_ca}
{Mashonkina}, L., {Korn}, A.~J., \& {Przybilla}, N. 2007, \aap, 461, 261,
  (Paper I)

\bibitem[{{Mashonkina} {et~al.}(2016){Mashonkina}, {Sitnova}, \&
  {Pakhomov}}]{Mashonkina_dnlte2016}
{Mashonkina}, L., {Sitnova}, T., \& {Pakhomov}, Y. 2016, Astronomy Letters, 42,
  606

\bibitem[{{Mel{\'e}ndez} {et~al.}(2007){Mel{\'e}ndez}, {Bautista}, \&
  {Badnell}}]{ca2_bautista}
{Mel{\'e}ndez}, M., {Bautista}, M.~A., \& {Badnell}, N.~R. 2007, \aap, 469,
  1203

\bibitem[{{Merle} {et~al.}(2011){Merle}, {Th{\'e}venin}, {Pichon}, \&
  {Bigot}}]{2011MNRAS.418..863M}
{Merle}, T., {Th{\'e}venin}, F., {Pichon}, B., \& {Bigot}, L. 2011, \mnras,
  418, 863

\bibitem[{{Mitrushchenkov} {et~al.}(2017){Mitrushchenkov}, {Guitou}, {Belyaev},
  {Yakovleva}, {Spielfiedel}, \& {Feautrier}}]{ca1_hydrogen}
{Mitrushchenkov}, A., {Guitou}, M., {Belyaev}, A.~K., {et~al.} 2017, J. Chem.
  Phys., 146, 014304

\bibitem[{{Moore}(1972)}]{1972mtai.book.....M}
{Moore}, C.~E. 1972, {A multiplet table of astrophysical interest - Pt.1: Table
  of multiplets - Pt.2: Finding list of all lines in the table of multiplets}

\bibitem[{Norris {et~al.}(2007)Norris, Christlieb, Korn, Eriksson, Bessell,
  Reimers, \& Wisotzki}]{Norrisetal:2007}
Norris, J., Christlieb, N., Korn, A., {et~al.} 2007, ApJ, 670, 774

\bibitem[{{N{\"o}rtersh{\"a}user} {et~al.}(1998){N{\"o}rtersh{\"a}user},
  {Blaum}, {Icker}, {M{\"u}ller}, {Schmitt}, {Wendt}, \&
  {Wiche}}]{1998EPJD....2...33N}
{N{\"o}rtersh{\"a}user}, W., {Blaum}, K., {Icker}, K., {et~al.} 1998, European
  Physical Journal D, 2, 33

\bibitem[{{Perryman} {et~al.}(2001){Perryman}, {de Boer}, {Gilmore}, {H{\o}g},
  {Lattanzi}, {Lindegren}, {Luri}, {Mignard}, {Pace}, \& {de
  Zeeuw}}]{2001A&A...369..339P}
{Perryman}, M.~A.~C., {de Boer}, K.~S., {Gilmore}, G., {et~al.} 2001, \aap,
  369, 339

\bibitem[{{Pfeiffer} {et~al.}(1998){Pfeiffer}, {Frank}, {Baumueller},
  {Fuhrmann}, \& {Gehren}}]{1998A&AS..130..381P}
{Pfeiffer}, M.~J., {Frank}, C., {Baumueller}, D., {Fuhrmann}, K., \& {Gehren},
  T. 1998, \aaps, 130, 381

\bibitem[{{Ryabchikova} {et~al.}(2015){Ryabchikova}, {Piskunov}, {Kurucz},
  {Stempels}, {Heiter}, {Pakhomov}, \& {Barklem}}]{2015PhyS...90e4005R}
{Ryabchikova}, T., {Piskunov}, N., {Kurucz}, R.~L., {et~al.} 2015, \physscr,
  90, 054005

\bibitem[{{Ryabchikova} {et~al.}(2016){Ryabchikova}, {Piskunov}, {Pakhomov},
  {Tsymbal}, {Titarenko}, {Sitnova}, {Alexeeva}, {Fossati}, \&
  {Mashonkina}}]{Ryabchikova2015}
{Ryabchikova}, T., {Piskunov}, N., {Pakhomov}, Y., {et~al.} 2016, \mnras, 456,
  1221

\bibitem[{{Rybicki} \& {Hummer}(1991)}]{rh91}
{Rybicki}, G.~B. \& {Hummer}, D.~G. 1991, \aap, 245, 171

\bibitem[{{Rybicki} \& {Hummer}(1992)}]{rh92}
{Rybicki}, G.~B. \& {Hummer}, D.~G. 1992, \aap, 262, 209

\bibitem[{{Samson} \& {Berrington}(2001)}]{2001ADNDT..77...87S}
{Samson}, A.~M. \& {Berrington}, K.~A. 2001, Atomic Data and Nuclear Data
  Tables, 77, 87

\bibitem[{{Scott} {et~al.}(2015){Scott}, {Grevesse}, {Asplund}, {Sauval},
  {Lind}, {Takeda}, {Collet}, {Trampedach}, \& {Hayek}}]{2015A&A...573A..25S}
{Scott}, P., {Grevesse}, N., {Asplund}, M., {et~al.} 2015, \aap, 573, A25

\bibitem[{{Seaton}(1962{\natexlab{a}})}]{1962PPS....79.1105S}
{Seaton}, M.~J. 1962{\natexlab{a}}, Proceedings of the Physical Society, 79,
  1105

\bibitem[{{Seaton}(1962{\natexlab{b}})}]{1962amp..conf..375S}
{Seaton}, M.~J. 1962{\natexlab{b}}, in Atomic and Molecular Processes, ed.
  D.~R. {Bates}, 375

\bibitem[{{Seaton} {et~al.}(1994){Seaton}, {Yan}, {Mihalas}, \&
  {Pradhan}}]{1994MNRAS.266..805S}
{Seaton}, M.~J., {Yan}, Y., {Mihalas}, D., \& {Pradhan}, A.~K. 1994, \mnras,
  266, 805

\bibitem[{{Sitnova} {et~al.}(2015){Sitnova}, {Zhao}, {Mashonkina}, {Chen},
  {Liu}, {Pakhomov}, {Tan}, {Bolte}, {Alexeeva}, {Grupp}, {Shi}, \&
  {Zhang}}]{2015ApJ...808..148S}
{Sitnova}, T., {Zhao}, G., {Mashonkina}, L., {et~al.} 2015, \apj, 808, 148

\bibitem[{{Smith}(1981)}]{1981A&A...103..351S}
{Smith}, G. 1981, \aap, 103, 351

\bibitem[{{Smith}(1988)}]{1988JPhB...21.2827S}
{Smith}, G. 1988, Journal of Physics B Atomic Molecular Physics, 21, 2827

\bibitem[{{Smith} \& {O'Neill}(1975)}]{1975A&A....38....1S}
{Smith}, G. \& {O'Neill}, J.~A. 1975, \aap, 38, 1

\bibitem[{{Smith} \& {Raggett}(1981)}]{1981JPhB...14.4015S}
{Smith}, G. \& {Raggett}, D.~S.~J. 1981, Journal of Physics B Atomic Molecular
  Physics, 14, 4015

\bibitem[{{Smith} \& {Gallagher}(1966)}]{1966PhRv..145...26S}
{Smith}, W.~W. \& {Gallagher}, A. 1966, Physical Review, 145, 26

\bibitem[{{Spite} {et~al.}(2012){Spite}, {Andrievsky}, {Spite}, {Caffau},
  {Korotin}, {Bonifacio}, {Ludwig}, {Fran{\c c}ois}, \&
  {Cayrel}}]{2012A&A...541A.143S}
{Spite}, M., {Andrievsky}, S.~M., {Spite}, F., {et~al.} 2012, \aap, 541, A143

\bibitem[{{Starkenburg} {et~al.}(2010){Starkenburg}, {Hill}, {Tolstoy},
  {Gonz{\'a}lez Hern{\'a}ndez}, {Irwin}, {Helmi}, {Battaglia}, {Jablonka},
  {Tafelmeyer}, {Shetrone}, {Venn}, \& {de Boer}}]{Starkenburg2010}
{Starkenburg}, E., {Hill}, V., {Tolstoy}, E., {et~al.} 2010, \aap, 513, A34

\bibitem[{{Steenbock} \& {Holweger}(1984)}]{Steenbock1984}
{Steenbock}, W. \& {Holweger}, H. 1984, \aap, 130, 319

\bibitem[{{Steinmetz} {et~al.}(2006){Steinmetz}, {Zwitter}, {Siebert},
  {Watson}, {Freeman}, {Munari}, {Campbell}, {Williams}, {Seabroke}, {Wyse},
  {Parker}, {Bienaym{\'e}}, {Roeser}, {Gibson}, {Gilmore}, {Grebel}, {Helmi},
  {Navarro}, {Burton}, {Cass}, {Dawe}, {Fiegert}, {Hartley}, {Russell},
  {Saunders}, {Enke}, {Bailin}, {Binney}, {Bland-Hawthorn}, {Boeche}, {Dehnen},
  {Eisenstein}, {Evans}, {Fiorucci}, {Fulbright}, {Gerhard}, {Jauregi}, {Kelz},
  {Mijovi{\'c}}, {Minchev}, {Parmentier}, {Pe{\~n}arrubia}, {Quillen}, {Read},
  {Ruchti}, {Scholz}, {Siviero}, {Smith}, {Sordo}, {Veltz}, {Vidrih}, {von
  Berlepsch}, {Boyle}, \& {Schilbach}}]{RAVE2006AJ....132.1645S}
{Steinmetz}, M., {Zwitter}, T., {Siebert}, A., {et~al.} 2006, \aj, 132, 1645

\bibitem[{{Theodosiou}(1989)}]{1989PhRvA..39.4880T}
{Theodosiou}, C.~E. 1989, \pra, 39, 4880

\bibitem[{{Watanabe} \& {Steenbock}(1985)}]{1985A&A...149...21W}
{Watanabe}, T. \& {Steenbock}, W. 1985, \aap, 149, 21

\bibitem[{{Zhao} {et~al.}(2016){Zhao}, {Mashonkina}, {Yan}, {Alexeeva},
  {Kobayashi}, {Pakhomov}, {Shi}, {Sitnova}, {Tan}, {Zhang}, {Zhang}, {Zhou},
  {Bolte}, {Chen}, {Li}, {Liu}, \& {Zhai}}]{lick_paperII}
{Zhao}, G., {Mashonkina}, L., {Yan}, H.~L., {et~al.} 2016, \apj, 833, 225

\end{thebibliography}
\bibliographystyle{aa}

\end{document}